\documentclass[11pt]{article}
\pdfoutput=1
\usepackage{geometry}                
\usepackage{graphicx}
\usepackage{amssymb}
\usepackage{amsmath}
\usepackage{epstopdf}
\usepackage{amscd}
\usepackage{mathtools}
\usepackage{slashed}
\usepackage{bbm}
\usepackage{upgreek}
\usepackage[utf8]{inputenc}
\usepackage[matrix,arrow]{xy}
\usepackage{arydshln}
\usepackage[utf8]{inputenc}

\usepackage{stolenstyle}
\usepackage{cjhebrew}

\usepackage[linktocpage=true]{hyperref}
\hypersetup{
colorlinks=true,
citecolor=blue,
linkcolor=blue,
urlcolor=blue}

\DeclareFontFamily{OT1}{pzc}{}
\DeclareFontShape{OT1}{pzc}{m}{it}{<-> s * [1.10] pzcmi7t}{}
\DeclareMathAlphabet{\mathpzc}{OT1}{pzc}{m}{it}

\newcommand{\overbar}[1]{\mkern 1.5mu\overline{\mkern-1.5mu#1\mkern-1.5mu}\mkern 1.5mu}

\def\Dbar{\overbar{D}}

\def\Fbar{\overbar{F}}

\def\Phibar{\overbar{\Phi}}

\def\taubar{\overbar{\tau}}
\def\psibar{\overbar{\psi}}
\def\gammabar{\overbar{\gamma}}
\def\UUbar{{\overbar{\mathcal{U}}}}
\def\QQbar{{\overbar{\mathcal{Q}}}}

\def\varepsbar{\overbar{\varepsilon}}

\def\etabar{\overbar{\eta}}

\def\Tr{ \, \textrm{Tr} \, }

\def\tm{{\widetilde{m}}}
\def\tn{{\widetilde{n}}}
\def\tp{\widetilde{p}}
\def\tq{\widetilde{q}}
\def\rtilde{{\widetilde{r}}}
\def\stilde{{\widetilde{s}}}
\def\ttilde{{\widetilde{t}}}
\def\utilde{{\widetilde{u}}}

\def\Im{ \, \mathrm{Im} }
\def\Re{ \, \mathrm{Re} }

\def\NN{\mathcal{N}}

\def\half{\frac{1}{2}}

\def\BB{\mathcal{B}}

\def\FF{\mathcal{F}}

\def\VV{\mathcal{V}}

\def\ZZ{\mathcal{Z}}

\def\DD{\mathcal{D}}

\def\JJ{\mathcal{J}}

\def\QQ{\mathcal{Q}}
\def\UU{\mathcal{U}}

\def\FFbar{{\overbar{\mathcal{F}}}}

\def\pd{\partial}

\def\vareps{\varepsilon}

\def\lambdabar{{\overbar{\lambda}}}
\def\xibar{{\overbar{\xi}}}

\def\psibar{{\overbar{\psi}}}

\def\muhat{{\hat{\mu}}}
\def\nuhat{{\hat{\nu}}}
\def\kappahat{{\hat{\kappa}}}
\def\xhat{{\hat{x}}}
\def\yhat{{\hat{y}}}
\def\yhatbar{{\overbar{\hat{y}}}}

\def\rhohat{{\hat{\rho}}}

\def\brho{{\boldsymbol{\rho}}}
\def\brhobar{{\overbar{\brho}}}
\def\bzeta{\boldsymbol{\upzeta}}
\def\bzetabar{\overbar{\boldsymbol{\upzeta}}}
\def\blambda{\boldsymbol{\uplambda}}
\def\blambdabar{\overbar{\boldsymbol{\uplambda}}}
\def\beps{\boldsymbol{\upepsilon}}
\def\bepsbar{\overbar{\boldsymbol{\upepsilon}}}
\def\bupxi{\boldsymbol{\upxi}}
\def\bupxibar{\overbar{\boldsymbol{\upxi}}}

\def\bpsi{\boldsymbol{\uppsi}}
\def\bpsibar{\overbar{\boldsymbol{\uppsi}}}
\def\bchi{\boldsymbol{\upchi}}
\def\bchibar{\overbar{\boldsymbol{\upchi}}}

\def\alphadot{{\dot{\alpha}}}
\def\betadot{{\dot{\beta}}}
\def\deltadot{{\dot{\delta}}}
\def\gammadot{{\dot{\gamma}}}

\def\sthetabar{{\overbar{\slashed{\theta}}}}

\def\sigmahat{{\hat{\sigma}}}
\def\ellhat{{\hat{\ell}}}
\def\ihat{{\hat{\imath}}}
\def\jhat{{\hat{\jmath}}}
\def\khat{{\hat{k}}}

\def\varepsbar{{\overbar{\varepsilon}}}
\def\etabar{{\overbar{\eta}}}

\def\lambdabar{{\overbar{\lambda}}}

\def\thetabar{{\overbar{\theta}}}

\def\ttheta{{\widetilde{\theta}}}
\def\talpha{{\widetilde{\alpha}}}
\def\tbeta{{\widetilde{\beta}}}
\def\tstheta{{\widetilde{\slashed{\theta}}}}
\def\tdelta{{\widetilde{\delta}}}
\def\tgamma{{\widetilde{\gamma}}}

\def\upphibar{{\overbar{\upphi}}}

\def\fbar{{\overbar{f}}}

\def\Phibar{{\overbar{\Phi}}}

\def\taubar{{\overbar{\tau}}}
\def\psibar{{\overbar{\psi}}}
\def\gammabar{{\overbar{\gamma}}}

\def\phibar{{\overbar{\phi}}}

\def\ttheta{{\widetilde{\theta}}}
\def\talpha{{\widetilde{\alpha}}}
\def\tbeta{{\widetilde{\beta}}}
\def\tstheta{{\widetilde{\slashed{\theta}}}}
\def\tdelta{{\widetilde{\delta}}}
\def\tgamma{{\widetilde{\gamma}}}

\def\stheta{{\slashed{\theta}}}
 
\def\ff{{\boldsymbol{f}}}
\def\ffbar{\overbar{{\boldsymbol{f}}}}
\def\bFF{\FF}

\def\bupzeta{{\boldsymbol{\upzeta}}}

\DeclareFontFamily{U}{rcjhbltx}{}
\DeclareFontShape{U}{rcjhbltx}{m}{n}{<->rcjhbltx}{}
\DeclareSymbolFont{hebrewletters}{U}{rcjhbltx}{m}{n}

\let\aleph\relax\let\beth\relax
\let\gimel\relax\let\daleth\relax

\DeclareMathSymbol{\aleph}{\mathord}{hebrewletters}{39}
\DeclareMathSymbol{\beth}{\mathord}{hebrewletters}{98}
\DeclareMathSymbol{\gimel}{\mathord}{hebrewletters}{103}
\DeclareMathSymbol{\daleth}{\mathord}{hebrewletters}{100}

\DeclareMathSymbol{\lamed}{\mathord}{hebrewletters}{108}
\DeclareMathSymbol{\mem}{\mathord}{hebrewletters}{109}
\DeclareMathSymbol{\ayin}{\mathord}{hebrewletters}{96}
\DeclareMathSymbol{\tsadi}{\mathord}{hebrewletters}{118}
\DeclareMathSymbol{\qof}{\mathord}{hebrewletters}{113}
\DeclareMathSymbol{\shin}{\mathord}{hebrewletters}{152}

\DeclareMathSymbol{\zain}{\mathord}{hebrewletters}{122}
\DeclareMathSymbol{\kafsof}{\mathord}{hebrewletters}{75}

\def\preprint{}

\title{Supersymmetry of the D3/D5 Defect Field Theory}
\author[a] {Sophia K.~Domokos}
\author[b] {and Andrew B.~Royston}

\abstract{Four-dimensional $\NN=4$ super Yang-Mills, with a codimension-one defect breaking half of the supersymmetry, arises as the field theory description of the D3/D5 intersection in the holographic limit. This is one of the earliest, most extensively studied, and commonly used systems in holography.  In this note we give the full R-symmetry-covariant supersymmetry variations for this system. We also provide the supercurrents and compute the algebra of the corresponding supercharges, obtaining the full set of central charges.  We show that magnetically charged finite-energy field configurations preserving half of the supersymmetry are solutions to a new form of the extended Bogomolny equations, in which the defect fields play the role of jumping data for the Nahm-like part of the equations.  In the appendices, we explain the connection between our results and the superspace-based formulations in the literature.}

\affiliation[a]{Department of Physics, New York Institute of Technology, 16 W. 61st Street, New York, NY 10023}
\affiliation[b]{ Department of Physics, Penn State Fayette, The Eberly Campus, 2201 University Drive, Lemont Furnace, PA 15456 }
\emailAdd{sdomokos@nyit.edu}
\emailAdd{abr84@psu.edu}

\begin{document}

\maketitle
\parskip 7pt

\section{Introduction}

Four-dimensional maximally supersymmetric Yang-Mills theory coupled to a codimension-one defect is one of the most throughly studied systems in the realms of integrability, AdS/CFT, and brane physics.  This system arises in string theory as the field theory on D3-branes intersected by D5-branes, which form a defect.  This brane intersection, and a generalization including NS5-branes, was used in studies of $\NN_{\rm 3d} = 4$ supersymmetric theories, where it provides an explanation for the connection between Coulomb branch vacua of such theories and monopole moduli spaces \cite{Hanany:1996ie,Cherkis:1997aa}.  It was also the focus of some of the first works to introduce fundamental flavors in AdS/CFT models and the first example of defect AdS/CFT \cite{Karch:2000gx, DeWolfe:2001pq}, where \cite{Erdmenger:2002ex} showed furthermore that the theory with a single defect is conformal. 

In the twenty-some years since these first works, a vast literature has explored this system as the starting point of many applications, from using integrability techniques \cite{deLeeuw:2015hxa, DeLeeuw:2018cal} and localization  \cite{Komatsu:2020sup} to extract CFT data, to the study of supersymmetric boundary conditions in $\NN_{\rm 4d} = 4$ supersymmetric Yang-Mills (sYM) \cite{Gaiotto:2008sa}, to models that break all supersymmetry and use the defect theory to approximate 2d condensed matter systems (see \cite{Jensen:2010ga} for seminal work).

Despite the extensive literature on this 4d-ambient plus 3d-defect theory, the complete supersymmetry variations under which it is invariant have not -- to our knowledge -- appeared in the literature.  These results could be useful for studying topologically twisted versions of an ambient plus defect theory along the lines of \cite{Witten:1988ze}, where R-symmetry plays a key role. The explicit form of the variations also plays an essential role in describing Bogomolny-Prasad-Sommerfield (BPS) solitons in this system.

We emphasize the difference between vacua, which have zero energy and preserve all supersymmetries, and solitons, which are local (but not global) minimizers of the energy and break some supersymmetry. Indeed, the theory in question features interesting vacua realized as solutions to the Nahm equations and Nahm-type equations \cite{Cherkis:2011ee,Gaiotto:2008sa}. This structure can be exhibited without recourse to the explicit form of the supersymmetry variations, for instance using the D- and F-term equations in a superfield-based formalism \cite{Cherkis:2011ee}.  For soliton states, however, the complete supersymmetry transformations of the ambient plus defect theory provide the most direct way to characterize the  spectrum of possible BPS solitons.\footnote{The quantum field-theoretic version of these statements is as follows.  The theory exhibits a parameter space of quantum vacua, with points in this space corresponding to solutions to the Nahm-type equations as discussed in \cite{Gaiotto:2008sa}.  This space is analogous to the Coulomb branch of vacua in $\NN_{\rm 4d} = 2$ supersymmetric theories.  Above each vacuum there is a Hilbert space of states, which includes as a subspace the Hilbert space of BPS states.  The classical finite-energy soliton configurations we discuss here would give rise to BPS states above these vacua upon semiclassical quantization.}

In this note, we write down the full R-symmetry-covariant supersymmetry variations of $\NN_{4d}=4$ sYM with a codimension-one defect of D5-brane type, preserving $\NN_{\rm 3d} = 4$ supersymmetry. We also give the supercurrents and the algebra of the supercharges following \cite{Witten:1978mh}, deriving central charges which measure both ambient and defect topological charges of field configurations.  While a complete analysis of BPS solitons is left for future work, we show here that magnetically charged solitons are described by solutions to a new form of the extended Bogomolny equations \cite{Kapustin:2006pk} in which the defect fields play the role of jumping data.  This system of equations, reviewed in Section \ref{sec:EBE}, is a set of generalized self-duality equations which has previously seen applications in physics-based constructions of Langlands duality \cite{Kapustin:2006pk} and knot invariants \cite{Witten:2011zz,Gaiotto:2011nm}.  This work shows they also play a key role in describing finite-energy solitons in the D3/D5 defect theory, and several lines for future development are suggested in the Conclusions.  

We also expect the soliton states of the D3/D5 defect theory discussed here to generate new insights, via holography, into the vacua and solitons of the dual gravitating theory explored in \cite{Domokos:2017wlb}. The present work is a necessary first step in that direction, which is currently under investigation.

In the interest of clarity and brevity, we suppress the details of all calculations, providing only explanations of key steps.\footnote{We are, however, happy to provide step-by-step notes upon request.} Despite being conceptually straightforward, constructing the variations in covariant form and demonstrating the invariance of the action were technically nontrivial. One reason for this is that much of the original literature works in superspace formalisms that obscure part of the full  $\mathrm{SU}(2)_{\rm V} \times \mathrm{SU}(2)_{\rm H}$ R-symmetry.    In the process of arriving at our results, we derived maps between the $\NN_{\rm{3d}}=2$ superfields of Erdmenger et al. \cite{Erdmenger:2002ex},  the formulation in terms of standard superspace coordinates  \cite{Wess:1992cp, Gates:1983nr}, and the R-symmetry-covariant component-field formulation. We include these in the appendices.  We also note that references \cite{Hanany:1996ie,Gaiotto:2008sd,Gaiotto:2008sa} identified the field content for this system with respect to $\NN_{\rm 3d} = 4$ supersymmetry representations.  While the general form that the supersymmetry transformations must take is clear from their work, getting the precise form was nontrivial due to the complicated nature of the system.

An outline of this paper is as follows.  Conventions are specified in Section \ref{subsec:conventions}.  We give the action and variations in Section \ref{sec:SUSY}.  In Section \ref{sec:boundary} we analyze the boundary terms from the variation and provide the supercurrents, which we use to compute the algebra of supercharges.  We give an application of these results in Section \ref{sec:EBE}, where we show that magnetically charged finite-energy BPS field configurations are solutions to the extended Bogomolny equations augmented with jumping data.  In Appendices \ref{app:SO4toSU2SU2} and \ref{app:superspace} we write down explicit maps from the fields we use to the common formulations of $\NN_{4d}=4$ sYM, and to previous constructions of the ambient plus defect theory based on superspace.

\section{Conventions}\label{subsec:conventions}
We work in a mostly plus metric convention, $\eta_{\mu\nu}=\text{diag}(-1,1,1,1)$.  The (3+1)-dimensional space housing the ambient sYM theory is parametrized by $x^\mu$ for $\mu = 0,...3$. The defect is localized in the $x^2 = y$ direction, and extended along $x^\muhat = (x^0, x^1, x^3) =: (x^{\hat{0}}, x^{\hat{1}}, x^{\hat{2}})$. 
 
 Our $\mathfrak{u}(N_{\rm c})$ Lie algebra conventions for the field strength and covariant derivative are $F_{\mu\nu} = 2 \pd_{[\mu} A_{\nu]} + [A_{\mu},A_{\nu}]$ and $\DD_{\mu} X = \pd_{\mu} X + [A_{\mu},X]$.  ``$\Tr$'' denotes a positive-definite Killing form on the Lie algebra normalized so that the generators satsify $\Tr(T^a T^b) = \half \delta^{ab}$.  Defect fields $q$ transform in the $\brho \oplus\brhobar$ representation, with $\brho$ denoting the fundemantal representation.  We take the Lie algebra generators to be represented by anti-Hermitian matrices, so that $\DD_\mu q = \pd_\mu q + A_{\brho\mu} q$.  This also implies, for example, that $(\DD_\mu q)^\dagger = \pd_\mu q^\dagger + q^\dagger (A_{\brho\mu})^\dagger = \pd_\mu q^\dagger - q^\dagger A_{\brho\mu}$. 
 
We denote 4d gamma matrices by $\gamma^\mu$ with $\gammabar = -i\gamma^0\gamma^1\gamma^2\gamma^3$, and we denote 3d gamma matrices by $\rho^\muhat$.  In both 3d and 4d, we define Dirac and charge conjugation of Dirac spinors $\psi$ in terms of intertwiners $A$ and $C$, with $A=A^\dagger$ and $C = - C^T$ as in \cite{Sohnius:1985qm}. 
\begin{align}
\text{Dirac conjugation:} \quad &\psibar := \psi^\dagger A\cr
\text{Charge conjugation:} \quad &\psi_c := C\psibar^T = C A^T \psi^\ast ~.
\end{align}
A Majorana spinor $\lambda$ satisfies $\lambda_c = \lambda$.  The intertwiners furnish similarity transformations between unitarily equivalent representations of the Clifford algebra $\gamma^\mu, (\gamma^\mu)^T, (\gamma^\mu)^\dagger$ such that
\begin{align}
A\gamma^\mu A^{-1} = (\gamma^\mu)^\dagger,\qquad C^{-1}(\gamma^\mu) C = - (\gamma^\mu)^T~.
\end{align}
Note that all of these relations hold for the 3d and 4d versions, respectively, of the intertwiners and gamma matrices $A_{(3)},~ C_{(3)},~\rho^\muhat$ and $A,~C,~\gamma^\mu$.

As we work almost entirely in terms of 3d and 4d Majorana spinors, a convenient basis for the $\gamma$ matrices --similar that of  \cite{DeWolfe:2001pq}-- is
\begin{align}\label{DFOlikebasis}
& \rho^{\hat{0}} = - \sigma^2~, \qquad \rho^{\hat{1}} = i \sigma^3~, \qquad \rho^{\hat{2}} = -i \sigma^1~, \cr
& \gamma^{\muhat} = \rho^{\muhat} \otimes \sigma^1 ~, \qquad \gamma^2 = i \mathbbm{1} \otimes \sigma^3~, \qquad \gammabar = \mathbbm{1} \otimes \sigma^2~.
\end{align}
Since $(\gamma^\mu)^\dag = - (\gamma^{\mu})^T$, we can take
\begin{equation}
C = - A = \sigma^2 \otimes \sigma^1 ~.
\end{equation}
We will interpret the first tensor factor of $C$ as the 3d charge conjugation matrix, so that
\begin{equation}\label{3dchargeconj}
C = C_{(3)} \otimes \sigma^1 = - A_{(3)} \otimes \sigma^1 ~, \qquad C_{(3)} = - A_{(3)} = \sigma^2~.
\end{equation}
The advantage of this basis is that  the 4d Majorana condition reduces block-diagonally to two 3d Majorana conditions and in each case $C A^T = \mathbbm{1}$, so each 4d Majorana spinor decomposes into a doublet 3d Majorana spinors:
\begin{equation}\label{4dMto3dM}
\textrm{Majorana spinor:} \quad \lambda = \left( \begin{array}{c} \blambda \\[1ex] \tilde{\blambda} \end{array} \right) ~, \qquad \textrm{with} \quad \blambda^\ast = \blambda~,\quad  \tilde{\blambda}^\ast = \tilde{\blambda}~.
\end{equation}
%

\section{Supersymmetry Variations}\label{sec:SUSY}

The field content of the D3/D5 defect field theory consists of 
\begin{align}
&\text{Ambient:} \qquad  X^{\rm{V}}_{\rtilde},~X^{\rm{H}}_r,~ \lambda_{\tm m},~A_\mu\cr
&\text{Defect:}\qquad  q^m, \bupzeta_{\tm} ~. \nonumber
\end{align}
The fields of $\NN_{\text{4d}}=4$ sYM  are grouped to transform under the 3d $\mathrm{SU}(2)_{\rm V}  \times \mathrm{SU}(2)_{\rm H}$ R-symmetry preserved in the presence of the defect, rather than the larger 4d $\mathrm{SU}(4)$ R-symmetry preserved by  $\NN_{\text{4d}}=4$ sYM without the defect.  $X^{\rm{V}}_{\rtilde},~X^{\rm{H}}_r$ are triplets of real scalars, and $\lambda_{\tm m}$ are 4d Majorana spinors transforming in the ${\bf (2,2)}$ of  $\mathrm{SU}(2)_{\rm V}  \times \mathrm{SU}(2)_{\rm H}$.  $A_\mu$ is the 4d gauge field, which decomposes into components $A_\muhat$ parallel to the defect and $A_2$ orthogonal to the defect.  The defect fields consist of complex scalars $q^m$ transforming in the fundamental of $\mathrm{SU}(2)_{\rm H}$ and trivially under $\mathrm{SU}(2)_{\rm V}$, and 3d Dirac fermions $\bzeta_\tm$ transforming in the fundamental of $\mathrm{SU}(2)_{\rm V}$ and  trivially under $\mathrm{SU}(2)_{\rm H}$.  The various index sets we utilize are summarized in Table \ref{indextable}. 

\begin{table}
\begin{center}
\begin{tabular}{| c | c | c |} \hline
index type & description & range \\ 
\hline \hline
$\mu,\nu,\rho,\sigma$ & 4d ambient spacetime & $0,1,2,3$ \\
$ i,j,k,\ell$  & 3d ambient space & $1,2,3$ \\
$\muhat,\nuhat,\rhohat,\sigmahat$ & 3d defect spacetime & $\hat{0},\hat{1},\hat{2}$  \\
$\ihat,\jhat,\khat,\ellhat$ & 2d defect space & $\hat{1},\hat{2}$ \\
$\alpha,\beta,\gamma,\delta$ & 4d positive chirality Weyl spinors & $1,2$ \\
$\alphadot,\betadot,\gammadot,\deltadot$ & 4d negative chirality Weyl spinors & $1,2$ \\
$\talpha,\tbeta\,\tgamma, \tdelta$ & 3d Dirac or Majorana spinors & $1,2$ \\
\hline
$a,b$ & Lie algebra & $1,\ldots, N_{c}^2$ \\
\hline
$r,s,t,u$ & $\mathrm{SU}(2)_{\rm H}$ triplet & $1,2,3$ \\
$m,n,p,q$ &  $\mathrm{SU}(2)_{\rm H}$ doublet  &  $1,2$ \\
$\rtilde,\stilde,\ttilde,\utilde$ & $\mathrm{SU}(2)_{\rm V}$ triplet &  $1,2,3$ \\
$\tm,\tn,\tp, \tq$  &  $\mathrm{SU}(2)_{\rm V}$ doublet  &  $1,2$ \\ 
$I,J,K,L$ & $\mathrm{SO}(4)$ quartet & $1,2,3,4$ \\
$A,B,C,D$ & $\NN_{\rm{4d}}=2$ $\mathrm{SU}(2)_{\rm{R}}$ doublet & $1,2$\\
\hline
\end{tabular}
\end{center}
\caption{Summary of index conventions and ranges. Note that indices on 4d Dirac and Majorana spinors are always suppressed.}
\label{indextable}
\end{table}

The $\mathrm{SU}(2)$ indices are contracted using a Euclidean metric $\delta_{rs}$ or $\delta_{\rtilde\stilde}$, while the $m$-type and $\tm$-type indices are raised and lowered using the Levi-Civit\`{a} symbols $\epsilon_{mn}, \epsilon_{\tm\tn}$ with $\epsilon_{12}= -1$ and $\epsilon^{12}= 1$. The fundamental of $\mathrm{SU}(2)$ is pseudo-real so complex conjugation raises/lowers the index.  Thus, in particular, $(q^m)^\dagger = q^\dagger_m$.

The spinors $\lambda_{\tm m}$ satisfy a Majorana condition that takes into account their R-symmetry transformation properties:
 \begin{align}\label{eq:modifiedMajoranaCond}
 \lambda_{\tm m}=\epsilon_{\tm \tn}\epsilon_{mn}C(\lambdabar^{n\tn})^T~.
 \end{align}
Here the transpose refers to the spinor space only.  This definition holds for their 3d components, $\blambda_{\tm m},\tilde{\blambda}_{\tm m}$ as well, with the appropriate charge conjugation operator $C_{(3)}$. In what follows, when we say that a spinor with two R-symmetry indices is Majorana, we mean that it is Majorana with respect to this condition.

\begin{figure}
\begin{center}
\includegraphics[scale=0.5]{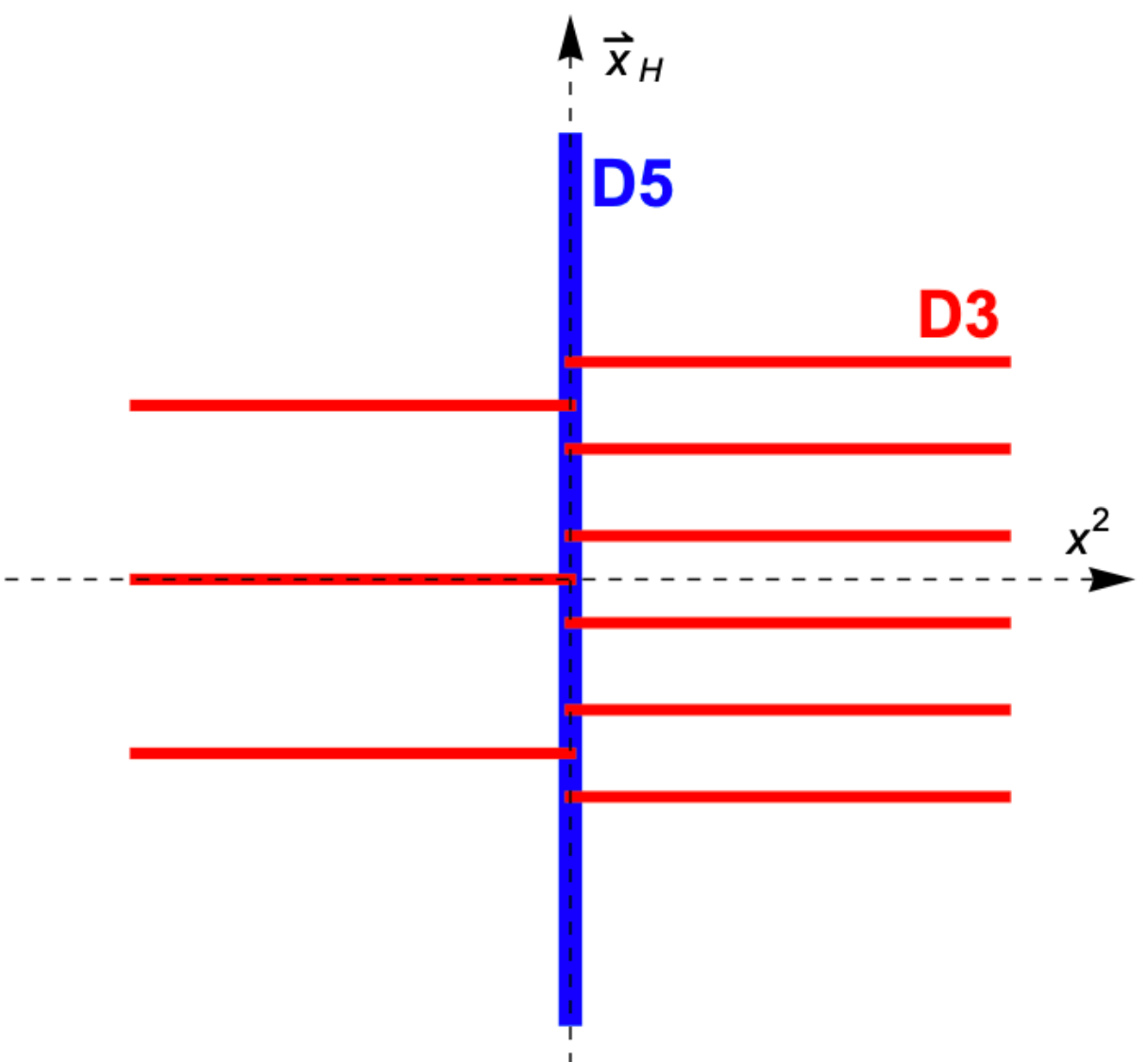}
\end{center}
\caption{Cartoon of the intersecting brane system.  The D5-brane stack sits at $x^2 = 0$.  The asymptotic position of the D3-branes in the directions labeled $\vec{x}_{\rm H}$ are determined by the values of the diagonal components of the triplet $X_{r}^{\rm H}$ as $x_2 \to \pm \infty$.}
\label{D3D5system}
\label{fig1}
\end{figure}

The transformation properties of the fields with respect to the gauge group are most easily understood from the intersecting D-brane picture.  Since D3-branes can end on D5-branes, there may be a different number of D3-branes to the left of the defect $(x^2 < 0)$, than to the right $(x^2 > 0)$.  See Figure \ref{fig1}.  If these numbers are $N_L$ and $N_R$, then the gauge group of the ambient theory for $x^2 < 0$ is $\mathrm{U}(N_L)$, and is $\mathrm{U}(N_R)$ for $x^2 > 0$.  For definiteness suppose $N_R \geq N_L$.  In what follows, we take the gauge group of the ambient theory to be $\mathrm{U}(N_{\rm  c})$ with $N_{\rm c} = \max(N_L,N_R)$, with the understanding that any adjoint-valued field $\Phi = \Phi^a T^a$ has $\Phi^a = 0$ for $x_2 < 0$ when $a > N_{L}^2$.  This amounts to choosing an embedding $\mathrm{U}(N_L) \subset \mathrm{U}(N_R)$, in which the left gauge group can be thought of as the upper-left block of the right gauge group in the defining representation.  

The set of possible boundary conditions on the ambient fields as $x^2 \to 0$, that preserve $\mathcal{N}_{\rm 3d} = 4$ supersymmetry, was specified in \cite{Gaiotto:2008sa}.  In brief, for theories corresponding to the D3/D5 intersection, the triplet $X_{r}^{\rm H}$ and the gauge field $A_2$ should have behavior consistent with a solution to Nahm's equation as $x^2 \to 0$, which might include a Nahm pole when $N_R > N_L + 1$.  A Nahm pole is specified by an embedding $\rho : \mathfrak{su}(2) \to \mathfrak{su}(N_R - N_L)$, and the possible embeddings are in turn determined by the number of coincident D5-branes present, as explained in \cite{Gaiotto:2008sa}.  If only a single D5-brane is present, then $\rho$ is the principal embedding.

The defect fields transform in the (anti-) fundamental of $\mathrm{U}(N_{\rm c,def})$, where $N_{\rm c,def} = \min(N_L,N_R)$, and  $q^m, q_{m}^\dag$ provide the jumping data for the upper left block of $X_{r}^{\rm H}$ at $x_2 = 0$.  If there are $N_f$ coincident D5-branes, then $q^m,\bzeta_{\tm}$ transform in the bifundamental $(\boldsymbol{N_{\rm c,def}},\overbar{\boldsymbol{N}_{\rm f}})$ of $\mathrm{U}(N_{\rm c,def})\times \mathrm{U}(N_f)$. Bilinears in the defect fields are always contracted to form $\mathrm{U}(N_f)$ singlets, and this structure is suppressed in what follows.  Thus, a quantity of the form $q^\dag q$ is a $\mathrm{U}(N_{\rm c,def})\times \mathrm{U}(N_f)$ singlet, while a quantity of the form $q^\dag T_{\brho}^a q$ transforms in the adjoint of $\mathrm{U}(N_{\rm c,def})$.  We will see below how the Nahm boundary conditions emerge from the perspectives of both supersymmetry and vanishing energy conditions.

Here again we emphasize that the Nahm-type configurations just discussed, including Nahm poles and nontrivial jumping data, are all vacuum configurations of the theory.  They preserve all of the $\NN_{\rm 3d} = 4$ supersymmetries and have zero energy with respect to the field theory Hamiltonian.  In particular the fields in such a configuration are independent of $x^\muhat$, such that these vacua preserve three-dimensional Poincar\'e symmetry.

The generalization of the above to multiple separated stacks of D5-branes and/or NS5-branes, leading to multiple parallel defects (of D5 and/or NS5 type) was described in \cite{Cherkis:2011ee}.  We will restrict ourselves here to the case of a single D3/D5-type defect.

With these preliminaries out of the way we can now write down the action in terms of these fields \cite{DeWolfe:2001pq}:
\begingroup
\allowdisplaybreaks
\begin{align}%
S =&~ S_{\rm amb} + S_{\rm def} ~, \qquad \textrm{with}  \cr
S_{\rm amb} =&~ \frac{1}{g_{\rm ym}^2} \int d^4 x \Tr \bigg\{ - \half F_{\mu\nu} F^{\mu\nu} - \frac{i}{2} \lambdabar^{m \tm} \gamma^{\mu} \DD_{\mu} \lambda_{\tm m} - \DD^{\mu} X_{\rtilde}^{\rm V} \DD_{\mu} X_{\rtilde}^{\rm V} - \DD^{\mu} X_{r}^{\rm H} \DD_{\mu} X_{r}^{\rm H}  \nonumber \\*
&~ \qquad   +  \frac{i}{2}  \lambdabar^{m \tm} [\lambda_{\tm n}, X_{r}^{\rm H}] (\sigma^r)^{n}_{~m} - \frac{1}{2} \lambdabar^{m \tm} (\sigma^{\rtilde})_{\tm}^{~\tn} \gammabar [\lambda_{\tn m}, X_{\rtilde}^{\rm V}] + \nonumber \\*
&~ \qquad - [X_{\rtilde}^{\rm V}, X_{s}^{\rm H}] [X_{\rtilde}^{\rm V}, X_{s}^{\rm H}] - \half  [X_{\rtilde}^{\rm V}, X_{\stilde}^{\rm V}] [X_{\rtilde}^{\rm V}, X_{\stilde}^{\rm V}]  - \half [X_{r}^{\rm H}, X_{s}^{\rm H}] [X_{r}^{\rm H}, X_{s}^{\rm H}] \bigg\} ~,   \label{eq:ambaction} \\
\intertext{and}
S_{\rm def} =&~  \frac{1}{g_{\rm ym}^2} \int d^3x \bigg\{ -\DD^{\muhat} q_{m}^\dag \DD_{\muhat} q^m - \frac{i}{2} ( \bzetabar^{\tm} \rho^{\muhat} \DD_{\muhat} \bzeta_{\tm} - (\DD_{\muhat} \bzetabar^{\tm}) \rho^{\muhat} \bzeta_{\tm} ) + \nonumber \\*
&~ \qquad \qquad + i \bzetabar^{\tm} (\sigma^{\rtilde})_{\tm}^{~\tn} (X_{\brho}^{\rm V})_{\rtilde} \bzeta_{\tn} + i q_{m}^\dag (\blambdabar_{\brho})^{m \tm} \bzeta_{\tm} + i \bzetabar^{\tm} (\blambda_{\brho})_{ \tm m} q^m + \nonumber \\*
&~ \qquad \qquad + q_{m}^\dag (X_{\brho}^{\rm V})_{\rtilde} (X_{\brho}^{\rm V})^{\rtilde} q^m + i q_{m}^\dag (\sigma^r)^{m}_{~n} \left( \DD_2 X^{\rm H}_r + \half \epsilon_{r}^{~st} [X_{s}^{\rm H}, X_{t}^{\rm H}] \right)_{\brho} q^n + \nonumber \\*
&~ \qquad \qquad  + \half \delta(0) ( q_{m}^\dag (\sigma^r)^{m}_{~n} T^{a}_{\brho} q^n ) ( q_{p}^\dag (\sigma^r)^{p}_{~q} T^{a}_{\brho} q^{q} ) \bigg\}  \label{eq:defaction}~,
\end{align}
\endgroup
where the 3d spinor $\blambda_{\tm m}$ is the piece of the 4d ambient field $\lambda_{\tm m}$ that interacts with the defect fields (defined in detail below). $T^a_\brho$ are the generators of $\mathfrak{u}(N_{\rm c,def})$ in representation $\brho$. 
The ambient action is the usual $\NN_{4d}=4$ sYM action, but  manifesting the  $\mathrm{SU}(2)_{\rm H}\times \mathrm{SU}(2)_{\rm V} \subset \mathrm{SU}(4)$ R-symmetry preserved by the defect. In Appendix \ref{app:SO4toSU2SU2} we describe the map between this form and a more standard form \cite{Sohnius:1985qm}.

While we have grouped the final term in \eqref{eq:defaction} with the defect action, this term proportional to $\delta(0)$ is in a separate category.  Terms of this type were first discussed in \cite{Sethi:1997zza, Kapustin:1998pb}.  The interpretation of the $\delta(0)$ factor as a UV regulator in the specific term in \eqref{eq:defaction} was described in \cite{DeWolfe:2001pq}.  It is an artifact of the low energy limit and would not be present in the full string field theory where branes have a finite thickness.  If one considers classical finite-energy field configurations, this term should be grouped with others to form a complete square; on the minimum energy configuration, the entire squared term vanishes.

The total action $S_{\rm amb} + S_{\rm def}$ is invariant under variations
\begingroup
\allowdisplaybreaks
\begin{align}\label{eq:SUSYvars}
\text{\underline{Ambient fields}}&\cr
\delta A_{\mu} =&~ \frac{i}{2} \varepsbar^{m \tm} \gamma_\mu \lambda_{\tm m}~, \qquad   \delta X_{\rtilde}^{\rm V} = \frac{1}{2} \varepsbar^{m \tm} (\sigma_{\rtilde})_{\tm}^{~\tn} \gammabar \lambda_{\tn m} ~, \qquad  \delta X_{r}^{\rm H} =  \frac{i}{2} \varepsbar^{m \tm} \lambda_{\tm n} (\sigma_r)^{n}_{~m} ~, \nonumber \\*
\delta \lambda_{\tm m} =&~  \frac{1}{2} \gamma^{\mu\nu} F_{\mu\nu} \vareps_{\tm m}  +i \gammabar \gamma^{\mu} \DD_{\mu} X_{\rtilde}^{\rm V} (\sigma^{\rtilde})_{\tm}^{~\tn} \vareps_{\tn m} + \gamma^{\mu} \DD_{\mu} X_{r}^{\rm H} \vareps_{\tm n} (\sigma^r)^{n}_{~m} + \nonumber \\*
&~   - \frac{i}{2}  [X_{\rtilde}^{\rm V}, X_{\stilde}^{\rm V}] \epsilon^{\rtilde\stilde\ttilde} (\sigma_{\ttilde})_{\tm}^{~\tn} \vareps_{\tn m} - i  [ X_{\rtilde}^{\rm V}, X_{s}^{\rm H}] (\sigma^{\rtilde})_{\tm}^{~\tn} \gammabar \vareps_{\tn n} (\sigma^s)^{n}_{~m} + \nonumber \\*
&~ + \frac{i}{2} [X_{r}^{\rm H}, X_{s}^{\rm H}] \epsilon^{rst} \vareps_{\tm n} (\sigma_t)^{n}_{~m} - i \delta(x^2) Q^{r} \vareps_{\tm n} (\sigma_r)^{n}_{~m}  \nonumber \\*
=:&~ \FFbar_{\tm m}^{n \tn}\vareps_{\tn n} ~,\nonumber \\*
\delta \lambdabar^{m \tm} =&~   - \half \varepsbar^{m \tm} \gamma^{\mu\nu} F_{\mu\nu} + i \varepsbar^{\tn m} \gamma^{\mu} \gammabar \DD_{\mu} X_{\rtilde}^{\rm V} (\sigma^{\rtilde})_{\tn}^{~\tm} + \varepsbar^{n \tm} \gamma^{\mu} \DD_{\mu} X_{r}^{\rm H} (\sigma^r)^{m}_{~n} + \nonumber \\*
&~ + \frac{i}{2}  \varepsbar^{m \tn} \epsilon^{\rtilde\stilde\ttilde} [X_{\rtilde}^{\rm V}, X_{\stilde}^{\rm V}] (\sigma_{\ttilde})_{\tn}^{~\tm} - i  \varepsbar^{n \tn} \gammabar [ X_{\rtilde}^{\rm V}, X_{s}^{\rm H}] (\sigma^{\rtilde})_{\tn}^{~\tm} (\sigma^s)^{m}_{~n} + \nonumber \\*
&~ - \frac{i}{2} \varepsbar^{n \tm} \epsilon^{rst} [X_{r}^{\rm H}, X_{s}^{\rm H}] (\sigma_t)^{m}_{~n} +i  \varepsbar^{n\tm} \delta(x^2) Q^{r} (\sigma_r)^{m}_{~n} \nonumber \\*
=:&~ \varepsbar^{n\tn}\FF^{m\tm}_{\tn n}~,\nonumber \\
\intertext{and}
\text{\underline{Defect fields}}&\cr
\delta q^{m} =&~  i\bepsbar^{m \tm} \bzeta_{\tm} ~,  \qquad   \delta q_{m}^\dag = -i \bzetabar^{\tm} \beps_{\tm  m}~,  \nonumber \\*
\delta \bzeta_{\tm} =&~  \rho^{\muhat} \beps_{\tm  m} \DD_{\muhat} q^m + X_{\rtilde}^{\rm V} (\sigma^{\rtilde})_{\tm}^{~\tn} \beps_{\tn m} q^m =: \ff^{m \tn}_{\tm} \beps_{\tn m}~, \nonumber \\*
\delta \bzetabar^{\tm} =&~ (\DD_{\muhat} q_{m}^\dag ) \bepsbar^{m \tm} \rho^{\muhat} - q_{m}^\dag \bepsbar^{m\tn} (\sigma^{\rtilde})_{\tn}^{~\tm} X_{\rtilde}^{\rm V} =: \bepsbar^{m\tn}\ffbar^{\tm}_{\tn m} ~,
\end{align}
\endgroup
with supersymmetry parameters $\vareps_{\tm m}$ and $\beps_{\tm  m}$, where $\vareps_{\tm m}$ is a 4d Majorana spinor, and $\beps_{\tm  m}$ a 3d Majorana spinor embedded in $\varepsilon_{\tm m}$ the same way that $\blambda_{\tm m}$ is embedded in $\lambda_{\tm m}$.   The quantity $Q^r$ in the ambient fermion variations is the combination
\begin{equation}\label{Qrdef}
Q^r := i(q_{m}^\dag (\sigma^r)^{m}_{~n} T_{\brho}^a q^n) T^a ~.
\end{equation}
As we are using an antihermitian representation for our generators $T_{\brho}^a$, $Q_{r}$ is an $\mathrm{SU}(2)_{\rm H}$ triplet valued in the real Lie algebra $\mathfrak{u}(N_{\rm c, def})$.  Note that we have defined $\ff^{m\tn}_{\tm}$ and $\FF^{m\tm}_{\tn n}$ to represent the groups of terms appearing in the  defect and ambient fermion variations, respectively.

Let us now fill in some details about these results and make a few comments.
\begin{itemize}
\item In the absence of the defect, the action would be 4d $\NN_{\text{4d}}=4$ sYM, invariant for any choice of $\vareps_{\tm m}$. The defect breaks the supersymmetry by half.  There is a map $P$ sending the 4d $\vareps_{\tm m}$ to a 3d Majorana spinor $\beps_{\tm  m}$:
\begin{align}\label{eq:P}
\beps_{\tm  m} := P\vareps_{\tm m}~.
\end{align}
The same map picks out the piece of the 4d ambient fermions $\lambda_{\tm m}$ that interacts with the defect:
\begin{align}
\blambda_{\tm m} := P\lambda_{\tm m}~.
\end{align}
For convenience, we also define $P'$, the map that acts on Dirac conjugates of these Majorana spinors using the 4d and 3d intertwiners $A$ and $A_{(3)}$:
\begin{align}\label{eq:PPrime}
\bepsbar^{m \tm} = \varepsbar^{m \tm} P' \quad \Rightarrow \quad P' := A^{-1} P^\dagger A_{(3)}~.
\end{align}
\item There are in principle many possible ways to embed the space of 3d Majorana spinors in the space of 4d Majorana spinors, equivalent to different choices of $P$. However, requiring that the action \eqref{eq:ambaction}, \eqref{eq:defaction} be invariant under the variations \eqref{eq:SUSYvars} severely restricts the possibilities. The necessary and sufficient conditions on $P$ are:
\begin{align}\label{eq:Pconditions}
& P (1 + i \gamma^2) \vareps_{\tm m} = 0~, \qquad P \gamma^{\muhat} \vareps_{\tm m} = 0~, \cr
& P \gammabar \gamma^2 \vareps_{\tm m} = 0~, \qquad P \gamma^{\muhat 2} \vareps_{\tm m} = 0 ~\cr
& ( i \rho^{\muhat} P + P\gammabar \gamma^{\muhat}) \vareps_{\tm m} = 0~, \qquad (\rho^{\muhat\nuhat} P - P \gamma^{\muhat\nuhat}) \vareps_{\tm m} = 0~.
\end{align}
The first two lines of \eqref{eq:Pconditions} impose conditions on $P$, while the final line imposes conditions on how the 3d Clifford algebra is embedded in the 4d Clifford algebra. The first two lines
are satisfied if half of the original $\NN_{\text{4d}}=4$ supersymmetries are eliminated via the projection
\begin{equation}\label{eq:3d4dprojection}
\frac{1}{2}(1+i\gamma^2)\vareps_{\tm m} =0~,
\end{equation}
and we take $P$ such that $P^\dag P$ gives the orthogonal projection:
\begin{align}\label{eq:PdagP}
P^\dagger P = \frac{1}{2} (1-i\gamma^2)~,
\end{align}
Note then that $\vareps_{\tm m} = P^\dag P \vareps_{\tm m} = P^\dag \beps_{\tm  m}$, so $P^\dag$ embeds $\beps_{\tm  m}$ into $\vareps_{\tm m}$.  The corresponding projection on the Dirac conjugate spinor is $\varepsbar^{m \tm} \half (1 - i \gamma^2) = 0$.  It follows from the definition of $P'$ that $P' P^{\prime \dag} = \half (1 + i \gamma^2)$ and so we similarly find that $\varepsbar^{m \tm} = \bepsbar^{m \tm} P^{\prime \dag}$.  We then also have $P^{\prime \dag} \vareps_{\tm m} = 0$.  In contrast, $\lambda_{\tm m}$ does not satisfy any projection.  Rather $\tilde{\blambda}_{\tm m} := P^{\prime \dag} \lambda_{\tm m}$ is an independent 3d Majorana spinor contained in $\lambda_{\tm m}$.  All of these statements take an extremely simple form in the basis \eqref{DFOlikebasis} where
 \begin{align}
 P^\dagger P =
\half (1 - i \gamma_2) = \half( \mathbbm{1}_4 + \mathbbm{1}_2 \otimes \sigma^3) = \left( \begin{array}{c c} \mathbbm{1} & 0 \\ 0 & 0 \end{array} \right)~,
\end{align}
and we can take
\begin{equation}\label{Pmatrix}
P = \left( \begin{array}{c c} \mathbbm{1} & 0 \end{array} \right) ~,\qquad P' = \left( \begin{array}{c} 0 \\ \mathbbm{1} \end{array} \right) ~.
\end{equation}

\item The ambient field variations are those of $\NN_{4d}=4$ -- which can be obtained by translating the Majorana form given in \cite{Sohnius:1985qm} to an $\mathrm{SU}(2)_{\rm V} \times \mathrm{SU}(2)_{\rm H}$ covariant form as described in Appendix \ref{app:SO4toSU2SU2} -- augmented by terms involving the combination of defect fields $Q^r$.  This modification is due to a contribution from the defect to the ambient auxiliary $D$-term equation of motion.  As the auxiliary $D$-field appears only in the ambient fermion variation, it is only this variation that is modified from its defect-free counterpart.
\item  Due to the defect-induced modification of the ambient fermion variation, the ambient action \eqref{eq:ambaction} now varies as
\begin{align}\label{ambientinflow}
g_{\rm ym}^2 \delta S_{\rm amb} =&~  \int d^4 x \Tr \bigg\{  -\varepsbar^{m \tm} \delta(x^2) Q^r (\sigma_{r})^{p}_{~m} \times \cr
&~ \qquad  \times  \bigg( - \gamma^{\mu} \DD_{\mu} \lambda_{\tm p} + [\lambda_{\tm n}, X_{s}^{\rm H}] (\sigma^{s})^{n}_{~p}  - i  (\sigma^{\stilde})_{\tm}^{~\tn} \gammabar [\lambda_{\tn p}, X_{\stilde}^{\rm V}]  \bigg) \bigg\} ~, \qquad  
\end{align}
up to total derivatives which will be discussed in the next section. These inflow terms \eqref{ambientinflow} cancel against those terms in the variation of the defect action that do not involve $\bzeta_{\tm}$.  Variations of terms in \eqref{eq:defaction} that contain the defect fermions $\bzeta_{\tm}$ cancel amongst themselves.
\item The variation of the $\delta(0)$ term in the action cancels against terms that result from evaluating the defect-induced modification of the $\lambda_{\tm m}$ variation on the defect.
\item We can see from the perspective of supersymmetry how the vacua described by solutions to Nahm's equations arise.  Noting that $\gamma^2 \vareps_{\tm m} = i \varepsilon_{\tm m}$, one sees that three of terms in the ambient fermion variation can be combined as
\begin{equation}
\delta \lambda_{\tm m} \supset i \left(  \DD_{2} X_{r}^{\rm H}  + \frac{1}{2} \epsilon_{r}^{~st} [X_{s}^{\rm H}, X_{t}^{\rm H}]  -  \delta(x^2) Q_{r} \right) (\sigma^r)^{n}_{~m} \vareps_{\tm n}~.
\end{equation}
Thus, by setting the quantity in parentheses to zero and choosing appropriate vacuum conditions on $A_{\muhat}, X_{\rtilde}^{\rm V}$, the fermion variation vanishes without imposing any restrictions on the supersymmetry parameter $\beps_{\tm m}$.  The quantity in parentheses is precisely the Nahm equation, with jumping data specified by $Q_r$.
\end{itemize}

\section{Boundary Terms, Supercurrents, and Supercharges}\label{sec:boundary}

In this section we describe the boundary terms that arise while applying the supersymmetry variations in the action above. We also write down the supercurrents, as well as the supercharges and the resulting algebra.

\subsection{Boundary Terms and Supercurrent}

Under the supersymmetry transformations \eqref{eq:SUSYvars}, with spacetime-varying supersymmetry parameters, the total ambient plus defect action transforms as
\begin{align}\label{BoundaryAndCurrent}
g_{\rm ym}^2 \delta (S_{\rm amb} + S_{\rm def}) =&~ \int d^4 x \left\{ (\pd_{\mu} \varepsbar^{m \tm}) \mathcal{J}_{\tm m}^{\mu} + \pd_{\mu} \mathcal{B}^{\mu} \right\}   ~.
\end{align}
The $\BB_\mu$'s encapsulate the boundary terms, while the $\JJ^{\mu}_{\tm m}$ are supercurrents.  We discuss each in turn.

The boundary terms in \eqref{BoundaryAndCurrent} arise from various integrations by parts carried out to check invariance of the action under supersymmetry.  We find that they can be expressed entirely in terms of the supersymmetry variations of the fields: 
\begin{align}\label{totalBB}
\int d^4 x \pd_{\mu} \mathcal{B}^{\mu} =&~ \int d^4 x  \pd_{\mu} \mathcal{B}_{\rm amb}^{\mu} + \int d^3 x \pd_{\muhat} \mathcal{B}_{\rm def}^{\muhat}  +  \frac{2}{g_{\rm ym}^2} \int d^4 x \pd_2 \Tr \left\{ \delta(x^2) Q^r \delta X_{r}^{\rm H}\right\} ~,
\end{align}
with
\begin{align}\label{boundarycurrents}
\mathcal{B}_{\rm amb}^{\mu} :=&~ \frac{1}{g_{\rm ym}^2} \Tr \bigg\{ - 2 F^{\mu\nu} \delta A_{\nu} - 2 (\DD^{\mu} X^{{\rm V}\rtilde}) \delta X_{\rtilde}^{\rm V} - 2 (\DD^{\mu} X^{{\rm H}r}) \delta X_{r}^{\rm H} - \frac{i}{2} \delta \lambdabar^{m \tm} \gamma^{\mu} \lambda_{\tm m} \bigg\} ~, \cr
\mathcal{B}_{\rm def}^{\muhat} :=&~   \frac{1}{g_{\rm ym}^2} \left\{- (\DD^{\muhat} q_{m}^\dag) \delta q^m - \frac{i}{2} (\delta \bzetabar^{\tm}) \rho^{\muhat} \bzeta_{\tm}  - (\delta q_{m}^\dag) \DD^{\muhat} q^m + \frac{i}{2} \bzetabar^{\tm} \rho^{\muhat} \delta \bzeta_{\tm} \right\} ~.\raisetag{20pt}
\end{align}
Here we separated out the $\pd_{2} \{ \delta(x^2) \Tr(Q^r \delta X_{r}^{\rm H})\}$ term because it is special and requires some comment.  This term is canceled by a piece from $\delta \lambdabar^{m \tm} \gamma^2 \lambda_{\tm m}$ in $\pd_2\BB_{\rm amb}^2$.  We have chosen to write \eqref{totalBB} in the form shown, however, to emphasize that it can be expressed directly in terms of the field variations.  Interestingly, \eqref{totalBB} is of the same form as the boundary terms one obtains from a general variation of the action when deriving equations of motion.  Based on this observation, our approach to the discussion of boundary conditions will be to first consider what boundary conditions are consistent with the variational principle such that \eqref{totalBB} vanishes for general variations.  Having identified such a set, we can then check that these conditions are consistent with supersymmetry.  Consistency means that a Dirichlet-type condition $\delta \Phi = 0$ should not be imposed independently if $\delta$ is a supersymmetry variation, but rather should arise from conditions on other fields that appear in this variation.

Requiring \eqref{totalBB} to vanish determines the boundary conditions we must impose on the fields at spatial infinity\footnote{Assuming appropriate Neumann or Dirichlet conditions are set up at initial and final times.} and at the defect ($x^2=0$). The boundary terms at spatial infinity vanish with appropriate fall-off condition on the fields. These conditions play an important role in determining the types of soliton states present in the system, and are better discussed in conjunction with the central charges (in Subsection \ref{subsec:supercharges}). 

The boundary at $x^2=0$, meanwhile, is trivial if the ambient fields are continuous across the defect. If, however, there are D3-branes ending on the D5-branes as in \cite{Hanany:1996ie,Gaiotto:2008sa}, the ambient fields can be discontinuous. In this case, a special choice of boundary conditions is needed to maintain supersymmetry in the presence of differing gauge groups on either side of the defect and/or when some D3-branes are broken across the D5-branes.  Recall from our discussion under Figure \ref{fig1}, that in the generic situation there are a number $N_{\rm c,def} = \min(N_L,N_R)$ of broken D3-branes together with $|N_L - N_R|$ additional D3-branes on one side.  The general boundary conditions for supersymmetric vacua discussed in \cite{Gaiotto:2008sa} are of Nahm type.  They include discontinuities in the $N_{\rm c, def} \times N_{\rm c,def}$ components of the Higgs field determined by jumping data constructed from the defect, together with a possible Nahm pole for the remaining components corresponding to the extra D3-branes on one side.  We now show how these boundary conditions are consistent with the vanishing of \eqref{totalBB}.

The case of the Nahm pole can be viewed as a Dirichlet condition on the $\NN_{\rm 3d} = 4$ hypermultiplet fields, where the leading behavior of the fields is held fixed.  The vector multiplet fields meanwhile satisfy Neumann conditions such that, overall, we have the boundary conditions
\begin{align}
&\delta X^{\rm{H}}_r \bigg|_{x^2=0} = 0 \ , \qquad \delta A_2 \bigg|_{x^2=0} = 0 \ , \qquad \tilde{\blambda}_{\tm m}\bigg|_{x^2=0}=0 \cr
& F^{2\nuhat} \bigg|_{x^2=0}=0 \ , \qquad  \DD_2 X^{\rm{V}}_\rtilde \bigg|_{x^2=0} = 0 \ .
\end{align}
These conditions guarantee that all boundary terms from a generic variation of the ambient action vanish.  Furthermore, because the supersymmetry variations $\delta X_{r}^{\rm H}$ and $\delta A_2$ only involve $\tilde{\blambda}$, this set of conditions is consistent with supersymmetry in the way described above.

The case of the Nahm pole should be contrasted with the $N_{\rm c, def} \times N_{\rm c,def}$ components of the fields, where one expects the triplet $X_{r}^{\rm H}$ to have a discontinuity at the defect.  This discontinuity should be related to the jumping data, but the left or right limiting value of $X_{r}^{\rm H}$ should be free to vary.  This expectation is nontrivially consistent with the form of the boundary terms we have found.  In this case we combine the final term of \eqref{totalBB} with the $\delta X_{r}^{\rm H} \DD_2 X_{r}^{\rm H}$ term contained in $\BB_{\rm amb}^2$ to form the combination
\begin{equation}\label{jumpbc}
\int d^4 \pd_\mu \BB^{\mu} \supset - \frac{2}{g_{\rm ym}^2} \int d^4 x \pd_2 \Tr \left\{ \delta X_{r}^{\rm H} ( \DD_2 X_{r}^{\rm H} - \delta(x^2) Q_r) \right\} ~.
\end{equation}  
This quantity will vanish provided that $\DD_2 X_{r}^{\rm H} - \delta(x^2) Q_r = 0$ in an infinitesimal neighborhood of the defect.  In other words, the discontinuity in the Higgs field triplet must be captured precisely by the jumping data.  The remaining boundary terms in \eqref{totalBB} at $x^2 = 0$ can then be made to vanish by demanding continuity for the other fields.  This ensures both consistency of the variational principle and preservation of supersymmetry.

Let us now return to \eqref{BoundaryAndCurrent} and discuss the supercurrent.  We find that it takes the form
\begin{align}\label{fullsupercurrent}
\mathcal{J}_{\tm m}^{\mu} = P^{\prime \dag} (\mathcal{J}_{\rm amb})_{\tm m} + \delta(x^2) \delta^\mu_{~ \muhat} \left((\mathcal{J}_{\rm def})^{\muhat}_{\tm m} + C_{(3)} (\overbar{\JJ}_{\rm def}^T)_{\tm m}^{\muhat} \right)
\end{align}
with
\begin{align}\label{43supercurrents}
(\mathcal{J}_{\rm amb})_{\tm m}^{\mu} :=&~  \frac{i}{g_{\rm ym}^2} \Tr \left\{ \FF_{\tm m}^{n \tn} \gamma^{\mu} \lambda_{\tn n} \right\} ~, \cr 
(\mathcal{J}_{\rm def})^{\muhat}_{\tm m} :=&~  \frac{i}{g_{\rm ym}^2} \ffbar_{\tm m}^{\tn} \rho^{\muhat} \bzeta_\tn ~, \qquad (\overbar{\mathcal{J}}_{\rm def})_{\muhat}^{m \tm} := -  \frac{i}{g_{\rm ym}^2} \bzetabar^\tn \rho_{\muhat} \ff_{\tn}^{m \tm} ~,
\end{align}
where $\bFF^{m \tm}_{n \tm}$ and $\ff^{m \tm}_{\tn}$ are the combinations of fields appearing in the ambient and defect fermion variations \eqref{eq:SUSYvars}. Note the appearance of the projector $P^{\prime \dag}$ in the first line, which expresses $\varepsbar^{m \tm}$ in terms of $\bepsbar^{m \tm}$.  In the absence of the defect, transformations parameterized by any $\vareps_{\tm m}$ are symmetries; with the defect, only the components of $\vareps_{\tm m}$ restricted by the projection \eqref{eq:PPrime} are. The current $\mathcal{J}_{\tm m}^{\mu}$ satisfies the Majorana condition \eqref{eq:modifiedMajoranaCond}, because $(\mathcal{J}_{\rm amb})_{\tm m}^{\mu}$ itself satisfies it, and the remaining contribution is a sum of $(\mathcal{J}_{\rm def})^{\muhat}_{\tm m}$ and its charge conjugate.

It's a nontrivial result that the supercurrent can be written entirely in terms of the field combinations appearing in the fermion variations alone. In particular, the defect-induced modification of the ambient fermion variation that appears in $\FF_{\tm m}^{n \tn}$ falls out of the variation of the defect action. As a consequence, the split of $\JJ^\mu_{\tm m}$ into $(\mathcal{J}_{\rm amb})_{\tm m}^{\mu}$ and $(\mathcal{J}_{\rm def})^{\muhat}_{\tm m}$ is a bit deceptive, since   $(\mathcal{J}_{\rm amb})_{\tm m}^{\mu}$  contains a term that localizes to the defect.

\subsection{Supercharges and the Supersymmetry Algebra}\label{subsec:supercharges}

With the supercurrent in hand, we have the field representation of the supercharges and we can compute their algebra. Of particular interest is the form of the central charges that appear there. 

The Noether (super)charges are spatial integrals of the time component of the current (at a fixed time):
\begin{align}\label{QNoether1}
\QQ_{\talpha \tm m} =&~ \int d^3 x (\JJ^0_{\tm m})_{\talpha}  \cr
=&~ \frac{i}{g_{\rm ym}^2} \int d^3 x \Tr \left\{ (P^{\prime \dag} \FF_{\tm m}^{n \tn} \gamma^0 \lambda_{\tn n})_\talpha \right\} + \cr &\qquad\qquad + \frac{i}{g_{\rm ym}^2} \int d^2 x \left( \ffbar^{\tn}_{\tm m} \rho^0 \bzeta_\tn - \epsilon_{\tn \tp} \epsilon_{mn} C_{(3)} ( \bzetabar^\tn \rho^0 \ff_{\tn}^{n \tp})^T \right)_{\talpha} ~,
\end{align}
where we have restored the 3d spinor index $\talpha$ for clarity. 

In order to determine the algebra, we compute the canonical momenta from the Lagrangian and impose canonical brackets between coordinates and momenta.\footnote{These are Dirac brackets for the fermions, since they satisfy second-class constraints.} We find that the bracket of two supercharges takes the form
\begin{align}\label{SUSYalgebra}
\{ \QQ_{\talpha \tm m} , \QQ_{\tbeta \tn n} \} =&~ -i \bigg\{ \epsilon_{\tm \tn} \epsilon_{mn} (\rho^{\muhat} C_{(3)})_{\talpha\tbeta} P_{\muhat}  + \left( (\sigma^r \epsilon)_{\tm \tn} \epsilon_{mn} \mathcal{Z}_{r}^{\rm V} + \epsilon_{\tm \tn} (\epsilon \sigma^r)_{mn} \mathcal{Z}_{r}^{\rm H} \right) (C_{(3)})_{\talpha\tbeta}   + \cr
&~ \qquad \qquad + (\sigma^r \epsilon)_{\tm \tn} (\epsilon \sigma^s)_{mn} (\mathcal{Z}^{\rm VH})_{rs}^{\hat{\imath}} (\rho_{\hat{\imath}} C_{(3)})_{\talpha\tbeta} \bigg\}~.  \raisetag{20pt}
\end{align}

The Hamiltonian $H = P^0 = -P_0$ and momenta are the $\NN_{\rm 4d}=4$ Hamiltonian and momenta, supplemented by contributions from the defect. The bosonic terms are
\begin{align}\label{HamP}
H =&~ H_{ \NN_{\rm 4d} = 4} + \frac{1}{g_{\rm ym}^2} \int d^2 x \bigg\{ \DD_{0} q_{p}^\dag \DD_{0} q^p + \DD_{\hat{\imath}} q_{p}^\dag \DD^{\hat{\imath}} q^p - q_{p}^\dag (X_{\brho r}^{\rm V})^2 q^p + \cr
&~ \qquad  \qquad \qquad \qquad \qquad  -i q_{p}^{\dag} (\sigma^r)^{p}_{~q} \left( \DD_2 X_{r}^{\rm H} + \half \epsilon_{r}^{~st} [X_{s}^{\rm H}, X_{t}^{\rm H}] \right)_{\brho} q^q + \cr
&~ \qquad \qquad \qquad \qquad \qquad  - \half \delta(0) ( q_{m}^\dag (\sigma^r)^{m}_{~n} T_{\brho}^a q^n) ( q_{p}^\dag (\sigma_r)^{p}_{~q} T_{\brho}^a q^{q}) \bigg\}~,\cr
P^{\hat{\imath}} =&~ P_{\NN_{\rm 4d} = 4}^{\hat{\imath}} - \frac{1}{g_{\rm ym}^2} \int d^2 x  \left\{ \DD_0 q_{p}^\dag D_{\hat{\imath}} q^p + \DD_{\hat{\imath}} q_{p}^\dag \DD_0 q^p \right\}
\end{align}
where the $\NN_{\rm 4d}=4$ versions are given by
\begin{align}\label{4DHamP}
H_{ \NN_{\rm 4d} = 4} =&~  \frac{1}{g_{\rm ym}^2} \int d^3 x \Tr \bigg\{ F_{i 0} F^{i}_{~0} + \half F^{ij} F_{ij}  + (\DD_{0} X_{\rtilde}^{\rm V})^2 + (\DD_{i} X_{\rtilde}^{\rm V})^2 +  (\DD_{0} X_{r}^{\rm H})^2 + (\DD_{i} X_{r}^{\rm H})^2 + \cr
& \qquad\qquad\qquad \qquad + \half [X_{\rtilde}^{\rm V}, X_{\stilde}^{\rm V}] ^2 +  [X_{\rtilde}^{\rm V}, X_{s}^{\rm H}]^2+ \half  [X_{r}^{\rm H}, X_{s}^{\rm H}]^2 \bigg\}~, \cr
P_{\NN_{\rm 4d} = 4}^{i} =&~  -\frac{1}{g_{\rm ym}^2} \int d^3x \Tr \bigg\{ 2 F_{j 0} F^{ji} + 2 (\DD_0 X_{\rtilde}^{\rm V}) (\DD^{i} X_{\rtilde}^{\rm V}) +2 (\DD_0 X_{r}^{\rm H}) (\DD^{i} X_{r}^{\rm H}) \bigg\} ~. \raisetag{20pt}
\end{align}
Note that the Hamiltonian \label{4DHamiltonianAndMom} contains the square
\begin{align}
H \supset &~ \frac{1}{g_{\rm ym}^2} \int d^3 x \Tr \left( \DD_2 X_{r}^{\rm H} + \half \epsilon_{r}^{~st} [X_{s}^{\rm H}, X_{t}^{\rm H}]  - \delta(x^2) Q_r  \right)^2 ~,
\end{align}
where one must use \eqref{Qrdef} and $\Tr(T^aT^b) = \half \delta^{ab}$.  This demonstrates the existence of the Nahm-type vacua mentioned earlier from the perspective of energy.  We emphasize that the Hamiltonian already contains this square without the need to complete the square by adding and subtracting a topological term.  The $\delta(0)$ term is essential for this property.

The central charges appearing in \eqref{SUSYalgebra} are 
\begin{align}\label{centralcharges}
\mathcal{Z}_{\rtilde}^{\rm V} =&~  \ZZ^{\rm V}_{{\rm el},\rtilde} + \frac{1}{g_{\rm ym}^2} \epsilon_{\rtilde}^{~\stilde \ttilde} \epsilon^{\hat{\imath}\hat{\jmath}} \int d^3x \pd_{\hat{\imath}} \Tr \left\{ X_{\stilde}^{\rm V} \DD_{\hat{\jmath}} X_{\ttilde}^{\rm V} \right\} ~, \cr
\mathcal{Z}_{r}^{\rm H} =&~ \ZZ^{\rm H}_{{\rm mag},r} +  \frac{1}{g_{\rm ym}^2} \epsilon_{r}^{~st} \epsilon^{\hat{\imath}\hat{\jmath}} \int d^3x \pd_{\hat{\imath}} \Tr \left\{ X_{s}^{\rm H} \DD_{\hat{\jmath}} X_{t}^{\rm H} \right\} + \frac{i}{g_{\rm ym}^2} \epsilon^{\hat{\imath}\hat{\jmath}} \int d^2 x \pd_{\hat{\imath}} \left( q_{p}^\dag (\sigma_r)^{p}_{~q} \DD_{\hat{\jmath}} q^q \right) ~, \cr
(\mathcal{Z}^{\rm VH})^{\hat{\imath}}_{\rtilde s} =&~ - \frac{2}{g_{\rm ym}^2} \epsilon^{\hat{\imath}\hat{\jmath}} \int d^3 x \pd_{\hat{\jmath}} \Tr \left\{ X_{\rtilde}^{\rm V} \left( \DD_2 X_{s}^{\rm H} - \delta(x^2) Q_s \right) \right\} ~,
\end{align}
where $\ZZ^{\rm V}_{{\rm el},\rtilde}$ and $\ZZ^{\rm H}_{{\rm mag},r}$ comprise the half of the $\NN_{\rm 4d}=4$ central charges that survive the projection to $\NN_{\rm 3d}=4$:
\begin{align}\label{N4charges}
&\ZZ^{\rm V}_{{\rm el},\rtilde} = - \frac{2}{g_{\rm ym}^2} \int d^3 x \pd^{i} \Tr \{ F_{i 0} X_{\rtilde}^{\rm V} \} ~, \qquad \ZZ^{\rm H}_{{\rm mag},r} = \frac{1}{g_{\rm ym}^2} \int d^3 x \epsilon^{ijk} \pd_{i} \Tr \{ F_{jk} X_{r}^{\rm H} \}  ~.
\end{align}
These measure 3d electric and magnetic flux, respectively, along  the $X^{\rm{V}}$ and $X^{\rm{H}}$  Higgs directions. 

Let us make some comments about the central charges:
\begin{itemize}
\item The $\mathcal{Z}^{\rm VH}$ term in \eqref{SUSYalgebra} explicitly breaks the Poincar\'{e} covariance of the algebra, and is not expected to contribute for finite-energy field configurations. It may, however, contribute if one considers BPS strings with finite energy per unit length.
 \item Standard fall-off conditions on the asymptotic $S^2$ of the 3d ambient space allow for finite-energy configurations with electric and magnetic charge. In addition, vortex-type boundary conditions on the asymptotic $S^1$ of the 2d defect space will yield finite contributions to $\ZZ_r^{\rm H}$ from the $q^\dagger \DD q$ term.
 \item The $X\DD X$-type terms that appear in both $\ZZ^{\rm V}_\rtilde$ and $\ZZ^{\rm H}_r$ are integrated over the 3d ambient space, but restrict not to an $S^2$ at infinity, but an ${\mathbbm R}_{x^2}\times S^1$. One can imagine a field configuration which has vortex-like asymptotics on the $S^1$, but which falls off along the ${\mathbbm R}_{x^2}$ as one moves away from the defect, thus resulting in finite energy. 
\item It is fascinating that the hypermultiplet central charge receives contributions from both 3d monopole charge, and 2d vortex charge. This strongly suggests that there exist BPS field configurations in which a non-abelian vortex string is stretched between a magnetic monopole in the ambient space and the defect.  In the next section we determine the system of equations such field configurations should solve so that they saturate the Bogomolny bound $H \geq |\mathcal{Z}^{\rm H}|$.
\end{itemize}

 \section{The Extended Bogomolny Equations with Jumping Data}\label{sec:EBE}

As noted in the Introduction, one of the motivations for constructing the supersymmetry variations in this system is to characterize finite energy BPS solitons. We take a first look at the case of magnetically charged half-BPS configurations here, deferring a complete
analysis of the BPS spectrum to future work.

\subsection{Supersymmetry Projection}

We are interested field configurations preserving a half of the $\NN_{\rm 3d} = 4$ supersymmetry.  Focusing on magnetically charged configurations, we make the following ansatz for the generators of the preserved supersymmetry:
\begin{align}\label{eq:BPSstateProjectors}
-i\gamma^2 \vareps_{\tm m} = \vareps_{\tm m} \qquad\text{and}\qquad \hat{n}^r(\sigma_r)^m_{\ n}\gamma^{123}\vareps_{\tm m} = \vareps_{\tm m} ~,
\end{align}
where $\gamma^{123}:=\gamma^1\gamma^2\gamma^3$ and $\hat{n}^r$ a unit vector picking out a direction in SU(2)$_{\rm H}$.
The first of these conditions is \eqref{eq:3d4dprojection}, the projection that eliminates half of the supersymmetry of $\NN_{\rm 4d}=4$ to yield the $\NN_{\rm 3d}=4$ of the ambient-plus-defect theory. The second projection reduces the $\NN_{\rm 3d} = 4$ supersymmetry by half, leaving a total of four independent generators.  The motivation for the form of this latter projection is that it must commute with the first one, it should preserve the Wigner little group of the $\mathrm{Spin}(1,2)$ Lorentz group for a massive particle, and it must break the $\mathrm{SU}(2)_{\rm H}$ symmetry if the corresponding solution is to carry a nontrivial central charge, $\mathcal{Z}_{r}^{\rm H}$.

We look for time-independent bosonic field configurations in static gauge ($A_0=0$), with $\vec{X}^{\rm V}=0$, that are left invariant under the variations \eqref{eq:SUSYvars} generated by those $\vareps_{\tm m}$ satisfying \eqref{eq:BPSstateProjectors}.  This amounts to the requirement that the fermion variations vanish. Under these restrictions on the fields, the fermion variations from \eqref{eq:SUSYvars} simplify to
\begin{align}\label{restrictedfermivar1}
\delta \lambda_{\tm m} &= \frac{1}{2}\gamma^{ij}F_{ij}\vareps_{\tm m}+\gamma^i \DD_i X^{\rm H}_r\vareps_{\tm n} (\sigma^r)^n_{\ m}+\frac{i}{2}[X^{\rm H}_r,X^{\rm H}_s]\epsilon^{rst}\vareps_{\tm n}(\sigma_t)^n_{\ m}-i\delta (x^2)Q^r \vareps_{\tm n} (\sigma_r)^n_{\ m}\cr
\delta \zeta_\tm &= \rho^\jhat \beps_{\tm m}\DD_\jhat q^m~. \raisetag{20pt}
\end{align}
It follows from \eqref{eq:BPSstateProjectors}, \eqref{eq:P}, and \eqref{eq:Pconditions} that
\begin{align}\label{BPSstateprojrels}
\gamma^{\hat{1}\hat{2}} \varepsilon_{\tm m} =&~ i \hat{n}_r \vareps_{\tm n} (\sigma^r)^{n}_{~m} , \cr
\gamma^{\ihat 2} \vareps_{\tm m} =&~ - \hat{n}_r\epsilon^{\ihat\jhat} \gamma_{\jhat} \vareps_{\tm n} (\sigma^r)^{n}_{~m} ~, \cr
\gamma^{\hat{1}} \vareps_{\tm m} =&~  i \hat{n}_s \gamma^{\hat{2}} \vareps_{\tm n} (\sigma^s)^{n}_{\tm} ~, \cr
\rho^{\hat{1}} \beps_{\tm m} =&~  i \hat{n}_s \rho^{\hat{2}} \beps_{\tm n} (\sigma^s)^{n}_{\tm} 
\end{align}
Recall that 2d spatial coordinates are related to 3d spatial coordinates as $(x^{\hat{1}},x^{\hat{2}}) := (x^1,x^3)$ along the defect and $x^2 := y$ orthogonal to the defect.   To minimize confusion, we will use $\ihat = \hat{1},\hat{2}$ and $y$ exclusively for spatial indices in the following.  Applying \eqref{BPSstateprojrels}, the variations \eqref{restrictedfermivar1} can be written in terms of a linearly independent set of spinors as follows:
\begin{align}
\delta\lambda_{\tm m} &= i \left[ F_{\hat{1}\hat{2}}\hat{n}_r +  \DD_y X^{\rm H}_r + \frac{1}{2} \epsilon_{r}^{~st}[X^{\rm H}_s,X^{\rm H}_t ]  -\delta (y)Q_r \right] (\sigma^r)^n_{\ m}\vareps_{\tm n}\cr
&\qquad + \left[ F_{y\hat{1}} \hat{n}_r +\DD_{\hat{2}} X^{\rm H}_r - \epsilon_{r}^{~st} \hat{n}_s \DD_{\hat{1}} X^{\rm H}_t \right] (\sigma^r)^n_{\ m} \gamma^{\hat{2}} \vareps_{\tm n} + \cr
&\qquad + i \left[ F_{\hat{2}y} + \DD_{\hat{1}} (\hat{n}^rX^{\rm H}_r) \right] \gamma^{\hat{2}}\vareps_{\tm m}\cr
\delta \zeta_\tm &= i \left[ \DD_{\hat{1}}q^n - i \hat{n}_s (\sigma^s)^n_{\ m}\DD_{\hat{2}}q^m \right] \rho^{\hat{1}}\beps_{\tm n}~.
\end{align}   

Demanding these variations vanish yields the BPS equations
\begin{align}\label{EBEnrform}
0 =&~ \hat{n}_r  F_{\hat{1}\hat{2}}+  \DD_y X^{\rm H}_r + \frac{1}{2} \epsilon_{r}^{~st}[X^{\rm H}_s,X^{\rm H}_t ]  -\delta (y) Q_r ~, \cr
0 =&~ \hat{n}_r F_{y\hat{1}}  +\DD_{\hat{2}} X^{\rm H}_r - \epsilon_{r}^{~st} \hat{n}_s \DD_{\hat{1}} X^{\rm H}_t ~, \cr
0 =&~ F_{\hat{2}y} + \hat{n}^r \DD_{\hat{1}} X^{\rm H}_r ~, \cr
0 =&~  \DD_{\hat{1}}q^n - i \hat{n}_r (\sigma^r)^n_{\ m}\DD_{\hat{2}}q^m ~,
\end{align}
where we recall that $Q_r = (i q_{p}^\dag (\sigma_r)^{p}_{~q} T_{\brho}^a q^q) T^a$.  If we choose\footnote{The physical meaning of this choice will be discussed in the next subsection.} $\hat{n}_r = (0,0,1)$, these equations are brought to a more familiar form:
\begin{align}\label{EBEjump}
F_{\hat{1}\hat{2}} +  \DD_y X^{\rm H}_3 + [X^{\rm H}_1,X^{\rm H}_2] =&~ \delta (y) Q_3 \cr
\DD_y X^{\rm H}_1 +[X^{\rm H}_2,X^{\rm H}_3]  =&~  \delta (y)  Q_1 \cr
\DD_y X^{\rm H}_2 + [X^{\rm H}_3,X^{\rm H}_1] =&~  \delta (y) Q_2   \cr
F_{y \hat{1}} + \DD_{\hat{2}} X^{\rm H}_3  =&~  0  \cr
F_{\hat{2} y} + \DD_{\hat{1}} X^{\rm H}_3 =&~ 0 \cr
\DD_{\hat{2}} X^{\rm H}_1 + \DD_{\hat{1}} X^{\rm H}_2 =&~   0  \cr
\DD_{\hat{2}} X^{\rm H}_2 - \DD_{\hat{1}} X^{\rm H}_1 =&~ 0  \cr
 \DD_{\hat{1}}q^m-i(\sigma^3)^m_{\ n}\DD_{\hat{2}}q^n =&~  0~. 
\end{align}
If we set $q^m = 0$ so that the last equation is trivial and the right-hand sides of the first three equations are zero, then these become the extended Bogomolny equations as given in \cite{Kapustin:2006pk}.

The extended Bogomolny equations originally appeared in the approach of \cite{Kapustin:2006pk} to Langlands duality, and they have been used more recently in a gauge theory construction of Khovonov homology for knot invariants \cite{Witten:2011zz,Gaiotto:2011nm}.  Given that the D3/D5 system plays a central role in the work of \cite{Witten:2011zz,Gaiotto:2011nm}, it is not surprising that we find the closely related set of equations, \eqref{EBEjump}, from our analysis.

The difference of \eqref{EBEjump} compared to previous formulations is in the defect fields $q^m$.  They appear on the right-hand sides of the first three equations exactly as jumping data would appear in Nahm's equations, and they satisfy a differential constraint on the defect given by the last equation:  They must be covariantly constant with respect to the connection $\delta^{m}_{~n} \DD_{\hat{1}} - i (\sigma^3)^{m}_{~n} \DD_{\hat{2}}$.

The reason the defect fields did not appear in previous formulations is that these references were interested in a different boundary value problem.  They were focused on studying the equations on a two-manifold times an interval or half-space in the $y$-direction.  In terms of the D-brane picture of Figure \ref{fig1} and the discussion underneath, all D3-branes end on the D5-brane from one side and there are no defect fields.  One considers boundary conditions of Nahm-pole type as $y \to 0$, possibly generalized to include 't Hooft defects at fixed points in the boundary, rather than discontinuity conditions through $y=0$ determined by jumping data. 

Thus, while there has been significant activity on the generalized Nahm-pole boundary value problem for the extended Bogomolny equations on a half space \cite{MR3289327,Mazzeo:2017qwz,MR4019897,MR4139043}, these equations appear not to have been studied on $\mathbbm{R}^3$.  One reason for this is that, without the jumping data on co-dimension one defects, one does not expect to be able to find interesting three-dimensionally localized solutions beyond ordinary monopole configurations for $(A_{\ihat},A_y,X_{3}^{\rm H})$.\footnote{By three-dimensionally localized, we mean solutions with moduli that represent the position of mobile solitons in three-dimensional space.}  Ordinary monopole configurations will solve the extended Bogomolny equations with $X_{1,2}^{\rm H} = 0$ since in that case the extended Bogomolny equations reduce to the ordinary Bogomolny equations.  The work of \cite{Gaiotto:2011nm} and intuition from the brane picture indicate that generic solutions to the extended Bogomolny equations look like nonlinear superpositions of a solution to Nahm's equations and a monopole configuration, at least in regions of moduli space corresponding to large separation of the monopoles from the defect plane.  Without jumping data, however, there are no non-constant solutions to Nahm's equations on all of $\mathbbm{R}$.  With the inclusion of the jumping data provided by the defect fields, we expect there do exist solutions representing, for example, smooth monopoles in the presence of the defect with different asymptotic Higgs field values for $\vec{X}_{r}^{\rm H}$ as $y \to \pm \infty$.  

Let us return to \eqref{EBEjump}.  In writing the right-hand sides of the first three equations, we are assuming an embedding $\mathfrak{u}(N_{\rm c,def}) \subset \mathfrak{u}(N_{\rm c})$, with $N_c = \max(N_L,N_R)$, where the $Q_r$ populate the upper left block of the larger matrix in a matrix representation.\footnote{Recall that the defect fields transform in the $(\mathbf{N_{\rm c,def}}, \overbar{\mathbf{N_f}})$, where $N_{\rm c,def} = \min(N_L,N_R)$, when there are unequal numbers $N_{L,R}$ of D3-branes on the left and right of the stack of $N_{\rm f}$ D5-branes.}  To fully specify the boundary value problem of interest, we must also specify the boundary conditions on the remaining blocks of the $N_{\rm c} \times N_{\rm c}$ matrix for all the ambient fields.  Here we take our cue from the vacuum conditions in \cite{Gaiotto:2008sa} and previous work on the extended Bogomolny equations with Nahm pole boundary conditions \cite{Witten:2011zz,Gaiotto:2011nm,MR3289327}.  The off-diagonal blocks of all ambient fields should vanish as $y \to 0$.  The lower right block of the Higgs triplet should have leading behavior as $y \to 0$ consistent with a supersymmetric vacuum of D3/D5 type, which may include a Nahm pole when $|N_L - N_R| \geq 2$.  The subleading behavior for the fields around the Nahm pole asymptotics was discussed in \cite{MR3289327}.

There are many exciting directions to pursue in studying the equations \eqref{EBEjump}, some of which we mention in the conclusions.  For now we will content ourselves with a computation of the energy of such solutions, assuming they exist.  We will show that this energy saturates the expected Bogomolny bound.

\subsection{Bogomolny Bound}

We start with the bosonic Hamiltonian \eqref{HamP} with \eqref{4DHamP}, restricted to static field configurations with $A_0 = 0 = \vec{X}^{\rm V}$:
\begin{align}
H =&~ \frac{1}{g_{\rm ym}^2} \int d^2 3 x \Tr \bigg\{ \half F^{ij} F_{ij} + (\DD_i X_{r}^{\rm H})^2 + \half [X_{r}^{\rm H}, X_{s}^{\rm H}]^2 \bigg\} + \cr
&~ + \frac{1}{g_{\rm ym}^2} \int d^2 x \bigg\{ \DD_{\ihat} q_{m}^\dag \DD^{\ihat} q^m  -i q_{m}^{\dag} (\sigma^r)^{m}_{~n} \left( \DD_y X_{r}^{\rm H} + \half \epsilon_{r}^{~st} [X_{s}^{\rm H}, X_{t}^{\rm H}] \right)_{\brho} q^n + \cr
&~ \qquad \qquad \qquad  \quad  - \half \delta(0) ( q_{m}^\dag (\sigma^r)^{m}_{~n} T_{\brho}^a q^n) ( q_{p}^\dag (\sigma_r)^{p}_{~q} T_{\brho}^a q^{q}) \bigg\} ~.
\end{align}
By completing squares and using $\Tr(T^a T^b) = \half \delta^{ab}$ and $\int d^2 x = \int d^3 x \delta(y)$, we can write this as
\begin{align}\label{BPSTtrick}
H =&~ \frac{1}{g_{\rm ym}^2} \int d^3 x \bigg\{ \half \left( \hat{n}_r F_{\hat{1}\hat{2}}^a + (\DD_y X_{r}^{\rm H})^a + \half \epsilon_{r}^{~st} [X_{s}^{\rm H}, X_{t}^{\rm H}]^a - \delta(y) Q_{r}^a \right)^2 + \cr
&~ \qquad \qquad+ \half \left(  \hat{n}_r F_{y \hat{1}}^a + (\DD_{\hat{2}} X_{r}^{\rm H})^a - \epsilon_{r}^{~st} \hat{n}_s (\DD_{\hat{1}} X_{t}^{\rm H})^a \right)^2 + \half \left( F_{\hat{2} y}^a + \hat{n}^r (\DD_{\hat{1}} X_{r}^{\rm H})^a \right)^2 \bigg\} + \cr
&~ + \frac{1}{g_{\rm ym}^2} \int d^2 x \left( \DD_{\hat{1}} q_{m}^\dag + i (\DD_{\hat{2}} q_{n}^\dag) (\sigma^r)^{n}_{~m} \hat{n}_r \right) \left( \DD_{\hat{1}} q^m - i \hat{n}_s (\sigma^r)^{m}_{~p} \DD_{\hat{2}} q^p \right) + \cr
&~ + \hat{n}^r \int d^3 x \Tr \bigg\{ \epsilon^{ijk} F_{ij} \DD_k X_{r}^{\rm H} - \epsilon_{r}^{~st} \left( \half F_{\hat{1}\hat{2}} [X_{s}^{\rm H}, X_{t}^{\rm H}] + \DD_{\hat{2}} X_{s}^{\rm H} \DD_{\hat{1}} X_{t}^{\rm H} \right) \bigg\}  + \cr
&~ + i \hat{n}^{r} \int d^2 x \left[ q_{m}^\dag (\sigma_r)^{m}_{~n} (F_{\hat{1}\hat{2}})_{\brho} q^n + (\DD_{\hat{1}} q_{m}^\dag) (\sigma_r)^{m}_{~n} \DD_{\hat{2}} q^n - (\DD_{\hat{2}} q_{m}^\dag) (\sigma_r)^{m}_{~n} \DD_{\hat{1}} q^n \right] ~,
\end{align}
for any unit vector $\hat{n}_r$.  The first three lines of this expression are a sum of squares of the left-hand sides of the equations \eqref{EBEnrform}.  They are positive definite.  The last two lines are total derivatives.  In fact, they are exactly $\hat{n}^r \mathcal{Z}_{r}^{\rm H}$, with the hypermultiplet central charges given in \eqref{centralcharges}.  Hence
\begin{equation}
H \geq \hat{n}^r \mathcal{Z}_{r}^{\rm H}~,
\end{equation}
with equality if and only if all of \eqref{EBEnrform} hold.  

Let us investigate the form of $\hat{n}^r \mathcal{Z}_{r}^{\rm H}$ on a solution to \eqref{EBEnrform}.  From \eqref{centralcharges} and \eqref{N4charges} we have
\begin{align}\label{ZrHagain}
\mathcal{Z}_{r}^{\rm H} =&~ \frac{1}{g_{\rm ym}^2} \int d^3 x \left[ \epsilon^{ijk} \pd_i \Tr(F_{jk} X_{r}^{\rm H}) + \epsilon_{r}^{~st} \epsilon^{\ihat\jhat} \pd_{\ihat} \Tr(X_{s}^{\rm H} \DD_{\jhat} X_{t}^{\rm H}) \right] + \cr
&~ +  \frac{i}{g_{\rm ym}^2} \int d^2 x \epsilon^{\ihat\jhat} \pd_{\ihat} \left( q_{m}^\dag (\sigma_r)^{m}_{~n} \DD_{\jhat} q^n \right)~.
\end{align}
From the second and third equation of \eqref{EBEnrform} it follows that
\begin{align}
\DD_{\hat{1}} X_{r}^{\rm H} = - \epsilon_{r}^{~st} \hat{n}_s \DD_{\hat{2}} X_{t}^{\rm H} - \hat{n}_r F_{\hat{2}y} ~, \qquad \DD_{\hat{2}} X_{r}^{\rm H} = \epsilon_{r}^{~st} \hat{n}_s \DD_{\hat{1}} X_{t}^{\rm H} - \hat{n}_r F_{y\hat{1}} ~.
\end{align}
One then finds
\begin{align}\label{ambientbc1}
\hat{n}_r \epsilon^{rst} \epsilon^{\ihat\jhat} \pd_{\ihat} \Tr \left( X_{s}^{\rm H} \DD_{\jhat} X_{t}^{\rm H} \right)  =&~ \pd_{\ihat} \Tr \left( X_{r}^{\rm H} \DD^{\ihat} X_{r}^{\rm H} \right) - \pd_{\ihat} \Tr \left( (\hat{n}^r X_{r}^{\rm H}) \DD^{\ihat} (\hat{n}^s X_{s}^{\rm H}) \right) \cr
=&~  \pd_{\ihat} \Tr \left( X_{r}^{\rm H} \DD^{\ihat} X_{r}^{\rm H} \right) - \pd_{\hat{1}} \Tr \left( (\hat{n}^r X_{r}^{\rm H}) F_{y \hat{2}} \right) - \pd_{\hat{2}} \Tr \left( (\hat{n}^r X_{r}^{\rm H}) F_{\hat{1}y} \right) . \cr
\end{align}
Using the first equation of \eqref{EBEnrform} and the Jacobi identity to eliminate $\epsilon^{rst} \Tr( X_{r}^{\rm H} [X_{s}^{\rm H}, X_{t}^{\rm H}])$, we can write
\begin{align}\label{ambientbc2}
\hat{n}^r \epsilon^{ijk} \pd_i \Tr \left( F_{jk} X_{r}^{\rm H}\right) =&~ 2 \pd_{\hat{1}} \Tr \left( (\hat{n}^r X_{r}^{\rm H}) F_{y \hat{2}} \right) + 2 \pd_{\hat{2}} \Tr \left( (\hat{n}^r X_{r}^{\rm H}) F_{\hat{1} y} \right) + \pd_y \Tr \left( (\hat{n}^r X_{r}^{\rm H}) F_{\hat{2} \hat{1}} \right) + \cr
&~ + \pd_y \Tr \left\{ X_{r}^{\rm H} \left( \DD_y X_{r}^{\rm H}  - \delta(y) Q_r \right) \right\}. 
\end{align}
Consider the last two terms in this expression, which are the same using the first equation of \eqref{EBEnrform}.  We claim they do not contribute boundary terms at $y = 0$.  This was demonstrated in \cite{MR3289327} for the Nahm pole boundary condition, where it was shown that the extended Bogomolny equations imply that $F_{\hat{1}\hat{2}}$ goes to zero as $y \to 0$ quickly enough to kill this term.\footnote{Reference \cite{MR3289327} considered the Kapustin-Witten equations on $\mathbbm{R}^3 \times \mathbbm{R}_+$ with Nahm pole boundary condition.  The extended Bogomolny equations are a dimensional reduction of this, assuming translation invariance in one of the $\mathbbm{R}^3$ factors.}  This term does not contribution boundary terms at $y = 0$ in the case of jumping data boundary conditions either.  This follows from our discussion around \eqref{jumpbc}, which we required to vanish for consistency of the variational principle.  Hence the only contributions to the integral of \eqref{ambientbc1} plus \eqref{ambientbc2} come from spatial infinity, and summing the two results we can write this as
\begin{align}\label{S2infinity}
& \int d^3 x \left\{  \hat{n}^r \epsilon^{ijk} \pd_i \Tr \left( F_{jk} X_{r}^{\rm H}\right) +  \hat{n}_r \epsilon^{rst} \epsilon^{\ihat\jhat} \pd_{\ihat} \Tr \left( X_{s}^{\rm H} \DD_{\jhat} X_{t}^{\rm H} \right) \right\}  = \cr
& \qquad \qquad = \int_{S_{\infty}^2} d\Omega \lim_{R \to \infty} R^2 \hat{R}^i \Tr \left\{ \half \epsilon_{ijk} F^{jk} (\hat{n}^r X_{r}^{\rm H}) + X_{r}^{\rm H} \DD_i X_{r}^{\rm H} \right\} ~.
\end{align}

This result shows that, on a solution to \eqref{EBEnrform}, the central charge \eqref{ZrHagain} can be expressed in terms of asymptotic data only.  For the ambient fields this data consists of the vacuum expectation values of the triplet $X_{\rm r}^{\rm H}$ at infinity in $\mathbbm{R}^3$ and the charges that determine the leading $1/R$ behavior for the gauge field and subleading $1/R$ behavior for the Higgs fields.  Finiteness of the energy requires these vevs and charges to be mutually commuting and time independent.  They might, however, take on different (gauge inequivalent) values on the two hemispheres at infinity corresponding to $y > 0$ and $y < 0$.  For the defect fields, $q^m$, this data consists of their asymptotic values on $S_{\infty}^1$, the intersection of the $y = 0$ plane and $S_{\infty}^2$.  This asymptotic data should be determined, up to gauge equivalence, by the complete set of data specifying the Nahm-type vacuum together with some topological charges.  We expect at least a pair of magnetic charges specifying the monopole content to the left and right of the defect, with the difference in these charges related to the subleading asymptotic behavior of the defect fields in the $x^{\hat{1}}$-$x^{\hat{2}}$ plane, but we leave the details for future work.

Having determined $\hat{n}^r \mathcal{Z}_{r}^{\rm H}$ in terms of this data, one must then vary $\hat{n}^r$ to achieve the strongest possible bound.  This will be when $\hat{n}$ is in the direction of $\vec{\mathcal{Z}}^{\rm H}$, leading to the bound
\begin{equation}
H \geq |\vec{\mathcal{Z}}^{\rm H}| ~.
\end{equation}
One can then rotate\footnote{An analogous $\mathrm{SO}(2)$ rotation is necessary in $\NN_{\rm 4d} = 2$ sYM theories when determining BPS dyon field configurations for a given point on the Coulomb branch of vacua.} the fields by an $\mathrm{SU}(2)_{\rm H}$ rotation that sends the direction $\hat{n} = \vec{\mathcal{Z}}^{\rm H}/|\vec{\mathcal{Z}}^{\rm H}|$ to the $\hat{k}$-direction.  The BPS equations \eqref{EBEnrform} will then take the form of the extended Bogomolny equations \eqref{EBEjump} in terms of these rotated fields.
%

\section{Conclusion}\label{sec:conclusion}
We have written down the full R-symmetry-covariant supersymmetry variations of $\NN_{4d}=4$ sYM with a non-abelian, codimension-one defect preserving $\NN_{\rm 3d} = 4$ supersymmetry. We also computed the supercurrents, supercharges, and the algebra. This fills a gap in the literature, and opens the door to studies of BPS states in a defect field theory with a holographic dual.

To illustrate the use of the supersymmetry variations in the study of BPS states, in Section \ref{sec:EBE} we took a first look at magnetically charged half-BPS field configurations in the defect theory.  With a natural ansatz for the projection on supersymmetry parameters $\varepsilon_{\tm m}$ that determine the preserved supersymmetry for such configurations, we arrived at an interesting set of BPS equations, \eqref{EBEnrform}.  We showed how these equations are an $\mathrm{SU}(2)_{\rm H}$-rotated form of the extended Bogomolny equations, augmented with jumping data constructed from the defect fields.  In principle, one could have arrived at this same system by a suitably clever manipulation of the Hamiltonian, as we showed in \eqref{BPSTtrick}.  Without our supersymmetry analysis, however, determining the whole family of equations parameterized by $\hat{n}_r$ in this fashion would have been challenging, and determining how solutions to these equations saturate a Bogomolny bound in terms of the central charges $\vec{\mathcal{Z}}^{\rm H}$ would have been impossible.

We stress that our work in Section \ref{sec:EBE} is only a preview, meant to illustrate the possible uses of the explicit supersymmetry variations and supersymmetry algebra obtained in this paper.  There is much further work to do in the study of BPS field configurations and BPS states in this system.  Some immediate questions and observations continuing from the analysis of Section \ref{sec:EBE} are as follows:
\begin{itemize}
\item One anticipates dyonic -- electrically charged -- versions of the half-BPS field configurations discussed above.  In semiclassical quantization these will have the interpretation of quarter-BPS states, and their energy will saturate a bound $H \geq |\vec{\mathcal{Z}}^{\rm H}| + |\vec{\mathcal{Z}}^{\rm V}|$.  They can be thought of as the incarnation of $\NN_{\rm 4d} = 4$ quarter-BPS dyons in the nontrivial Nahm-type vacua of the defect theory.  Their field configurations will have a nontrivial electric field and $\vec{X}^{\rm V}$ triplet, in addition to the magnetic field and $\vec{X}^{\rm H}$ triplet, and will satisfy an enhanced set of BPS equations.
\item  Can one find an explicit solution for the simplest nontrivial finite-energy configuration involving monopole charge?  One could look for a cylindrically symmetric solution representing a single $\mathrm{su}(2)$ monopole in the presence of the defect with nontrivial jumping data such that the asymptotic Higgs vevs are different as $y \to \pm \infty$.  Exact solutions typically provide invaluable insight so, while obtaining such solutions can be challenging, it is certainly worth pursuing.
\item  What are the precise asymptotic data needed to determine the central charge, $\vec{\mathcal{Z}}^{\rm H}$, and what is the physical interpretation of this data?  We indicated how this data should appear in the asymptotic expansion of the fields under \eqref{S2infinity}.  What conditions on the data must hold to ensure existence of solutions to \eqref{EBEjump}?  
\item Given such data, what is the moduli space of solutions to \eqref{EBEjump}?  The answers to these questions in the case of the Nahm pole boundary condition, generalized to include 't Hooft defects in the boundary, were conjectured in \cite{Witten:2011zz,Gaiotto:2011nm} and proven in \cite{MR4019897,MR4139043}.
\item The moduli space of solutions is especially interesting with respect to the analysis of Section \ref{sec:EBE} since it represents a moduli space for finite-energy solitons.  One could use it, for example, to consider collective coordinate dynamics and quantization for these solitons.  This would be an essential step in the semiclassical quantization of the corresponding BPS states.  These moduli spaces should be fibered over the space of vacua, which are themselves moduli spaces of solutions to Nahm's equations.  One anticipates interesting wall-crossing phenomena for BPS states as the vacuum parameters are varied, and we expect that semiclassical quantization will provide a fascinating perspective on this phenomena, as it did in $\mathcal{N}_{\rm 4d} = 2$ theories \cite{Moore:2015szp,Moore:2015qyu,Brennan:2016znk,Brennan:2018ura}.\footnote{A connection between the Kapustin-Witten equations and wall crossing for BPS states in a class of 4d-theories with co-dimension one defects was recently described in the talk \cite{MooreTalk}.} 
\end{itemize}


{\bf Acknowledgements:}  The work of ABR is supported by NSF grant number PHY-2112781.  SKD's work is supported by NSF grant number PHY-2014025. We thank Daniel Brennan, Siqi He, Ilarion Melnikov, Greg Moore, and Edward Witten for helpful correspondence.



\appendix

\section{$\mathrm{SO}(4)$ to $\mathrm{SU}(2)_{\mathrm{V}}\times \mathrm{SU}(2)_{\mathrm{H}}$ for $\NN_{\rm 4d}=4$ sYM}\label{app:SO4toSU2SU2}

 Here we describe the map between $\NN_{\rm 4d}=4$ sYM manifesting $\mathrm{SO}(4)$ R-symmetry to the $\mathrm{SU}(2)_{\mathrm{V}}\times \mathrm{SU}(2)_{\mathrm{H}}$ version we use in this work. We  begin with the $\NN_{\text{4d}}=4$ sYM action written in manifestly $\mathrm{SO}(4)$-invariant form:
\begin{align}\label{N4d4MajoranaL}
g_{\rm ym}^2 \mathcal{L} =&~ \Tr \bigg\{ - \half F_{\mu\nu} F^{\mu\nu} - i \lambdabar_I \gamma^\mu \DD_{\mu} \lambda^I - \frac{1}{4} \DD_\mu A_{IJ} \DD^{\mu} A^{IJ} - \frac{1}{4} \DD_{\mu} B_{IJ} \DD^{\mu} B^{IJ} + \cr
&~ \qquad - \lambdabar_I [ \lambda_J, A^{IJ}] - i \lambdabar_I \overbar{\gamma} [\lambda_J, B^{IJ}] - \frac{1}{16} [A_{IJ}, B_{KL} ] [A^{IJ}, B^{KL}] + \cr
&~ \qquad - \frac{1}{32} [A_{IJ}, A_{KL}] [A^{IJ}, A^{KL}] - \frac{1}{32} [B_{IJ} , B_{KL}] [ B^{IJ}, B^{KL}] \bigg\}~,
\end{align}
Here $\lambda^I$ are a quartet of Majorana spinors, and $A_{IJ}$ and $B_{IJ}$ are real, self-dual and anti self-dual matrices of scalars, respectively.  
 This is simply equation (13.4) of Sohnius's classic review \cite{Sohnius:1985qm} rewritten in terms of our spacetime and Lie algebra conventions. We also use $\overbar{\gamma}=-i\gamma^0\gamma^1\gamma^2\gamma^3$ while \cite{Sohnius:1985qm} chooses $\gamma_5 =\gamma^0\gamma^1\gamma^2\gamma^3=i\overbar{\gamma}$.  We use  $I,J= 1,\dots, 4$ for the $\mathrm{SO}(4)$ R-symmetry index, instead of \cite{Sohnius:1985qm}'s $i,j$. $I,J$ are raised and lowered with $\delta_{IJ}$.  
 
 Applying the following field redefinitions yields the ambient action \eqref{eq:ambaction}:
 \begin{align}
\lambda_I =&~ \frac{i}{2} (\tau_{I})^{m \tm} \lambda_{\tm m} \cr
\lambdabar_I =& ~-\frac{i}{2} \lambdabar^{m \tm} (\taubar_I)_{\tm m}\cr
A_{IJ} =&~  \left( \begin{array}{c c c c} 0 & X_{3}^{\rm H} & - X_{2}^{\rm H} & X_{1}^{\rm H} \\[1ex] -X_{3}^{\rm H} & 0 & X_{1}^{\rm H} & X_{2}^{\rm H} \\[1ex] X_{2}^{\rm H} & - X_{1}^{\rm H} & 0 & X_{3}^{\rm H} \\[1ex] - X_{1}^{\rm H} & - X_{2}^{\rm H} & - X_{3}^{\rm H} & 0 \end{array} \right) =  \vec{\eta}_{IJ} \cdot \vec{X}^{\rm H}~,  \cr
B_{IJ} =&~ \left( \begin{array}{c c c c} 0 & -X_{3}^{\rm V} & X_{2}^{\rm V} & X_{1}^{\rm V} \\[1ex] X_{3}^{\rm V} & 0 & -X_{1}^{\rm V} & X_{2}^{\rm V} \\[1ex] -X_{2}^{\rm V} &  X_{1}^{\rm V} & 0 & X_{3}^{\rm V} \\[1ex] - X_{1}^{\rm V} & - X_{2}^{\rm V} & - X_{3}^{\rm V} & 0 \end{array} \right) = - \vec{\etabar}_{IJ} \cdot \vec{X}^{\rm V} ~,
 \end{align}
where $\tau_I=(\vec{\sigma},-i \mathbbm{1} )$ and $\taubar_I=(\vec{\sigma},i \mathbbm{1} )$ are Euclidean signature sigma matrices, and ($\eta^r_{IJ}$,  $\etabar^\rtilde_{IJ}$) are the 't Hooft symbols. In order to rewrite \eqref{N4d4MajoranaL}  as \eqref{eq:ambaction} one must  make use of the identities
\begin{align}
(\eta^{s})^{IJ} (\eta_{r})_{IJ} =& ~ 4 \delta^{s}_{~r}\cr
\taubar^{I}_{\tm m} \eta^{r}_{IJ} (\tau^J)^{n \tn} =&~  -2i (\sigma^r)^{n}_{~m} \delta_{\tm}^{~\tn}    ~, \cr
\taubar^{I}_{\tm m} \etabar^{\rtilde}_{IJ} (\tau^J)^{n \tn} =&~  2 i (\sigma^{\rtilde})_{\tm}^{~\tn} \delta_{m}^{~n}  ~, \cr
\taubar_{\tm m}^I \eta^{[r}_{IJ} (\eta^{s]})^{JK} \tau_{K}^{n \tn} =&~ 2i \epsilon^{rst} (\sigma_t)^{n}_{~m} \delta_{\tm}^{~\tn}    ~, \cr
\taubar_{\tm m}^I \etabar^{[\rtilde}_{IJ} (\etabar^{\stilde]})^{JK} \tau_{K}^{n \tn} =&~ -2 i \epsilon^{\rtilde \stilde \ttilde} (\sigma_{\ttilde})_{\tm}^{~\tn} \delta_{m}^{~n}  ~, \cr
\taubar_{\tm m}^I ( \eta^{r}_{IJ} (\etabar^{\stilde})^{JK} + \etabar^{\stilde}_{IJ} (\eta^r)^{JK} ) \tau_{K}^{n \tn} =&~ 4 (\sigma^\stilde)_{\tm}^{~\tn} (\sigma^r)^{n}_{~m} ~.
\end{align}
%
%

\section{Superspace Formulations and Maps to Our Conventions}\label{app:superspace}
Both of the early works detailing the holography of the D3/D5 system  \cite{DeWolfe:2001pq, Erdmenger:2002ex}  start with an action in superspace which manifests a subset of supersymmetries but obscures the rest. These formulations do make it easier to write down the action and to show conformal invariance at the quantum level. However, it is nontrivial to rearrange the superfields' components to reflect the defect-plus-ambient theory's full $\mathrm{SU}(2)\times \mathrm{SU}(2)$ R-symmetry.  

De Wolfe et al. \cite{DeWolfe:2001pq} work initially in $\NN=1$ superspace, where the fields are grouped into a $\NN_{\rm{4d}}=1$ vector multiplet, a triplet of adjoint-valued chiral multiplets, and two $\NN_{\rm{3d}}=1$ complex multiplets transforming in the (anti)fundamental. They quickly move to component form, however, with which they are able to infer an R-symmetry-covariant formulation of the action. 

Erdmenger et al. \cite{Erdmenger:2002ex} work in $\NN=2$ superspace,  where the ambient fields are initially grouped into an $\NN_{\rm{4d}}=2$ vector multiplet and a $\NN_{\rm{4d}}=2$ adjoint-valued hypermultiplet, which they then decompose into a  $\NN_{\rm{3d}}=2$ vector and triplet of hypermultiplets (or actually a family of these, parameterized by the transverse coordinate $x^2$). The defect fields, meanwhile, consist of a pair of $\NN_{\rm{3d}}=2$ hypermultiplets transforming in the fundamental and anti-fundamental. The decomposition of the $\NN_{\rm{4d}}=2$  vector multiplet into $\NN_{\rm{3d}}=2$ multiplets has a nice superfield presentation for an abelian gauge group. 

For non-abelian gauge groups, however, the situation is more subtle: there does not appear to be an analogous gauge-covariant decomposition\footnote{At least as far we lilliputians can tell.}, a point which is not (to our knowledge) directly addressed in the literature.

For readers who might need to navigate between different formulations, we provide the map between the $\NN_{\rm{3d}}=2$ superfields of Erdmenger et al. \cite{Erdmenger:2002ex}, and the formulation in terms of standard superspace coordinates \`{a} la \cite{Wess:1992cp, Gates:1983nr}. This was done, in the abelian setting, by \cite{Erdmenger:2002ex}. We found the extension to the non-abelian case to be nontrivial. We also provide the map from the extended superspace formulation of \cite{Wess:1992cp, Gates:1983nr} to our $\mathrm{SU}(2)\times \mathrm{SU}(2)$ R-symmetry-covariant treatment (similar to \cite{DeWolfe:2001pq}'s). Our focus here will be entirely on ambient fields: the defect is manifestly 3d, so there is no map to be performed.

Extended $\NN_{\rm{4d}}=2$ superspace is parametrized by Grassmann-valued coordinates $\theta^A$ for $A=1,2$, where each $\theta^A$ is a Weyl spinor in 4d. We use the following conventions for complex conjugation and the raising and lowering of superspace indices:
\begin{equation}\label{Rspaceconjugate}
(\theta_{A\alpha})^\ast = \thetabar_{\alphadot}^A ~, \qquad (\theta_{A}^\alpha)^\ast = \thetabar^{A \alphadot}~, \qquad  (\theta^{A}_{\alpha})^\ast = \thetabar_{A \alphadot} ~, \qquad (\theta^{A \alpha})^\ast = \thetabar_{A}^{\alphadot} ~,
\end{equation}
and
\begin{equation}\label{Rspacetranspose}
\theta^A = \epsilon^{AB} \theta_B ~, \qquad \theta_A = \epsilon_{AB} \theta^B~, \qquad \thetabar^A = - \epsilon^{AB} \thetabar_B ~, \qquad \thetabar_A = - \epsilon_{AB} \thetabar^B ~.
\end{equation}
We denote these coordinates Wess-Bagger (WB) coordinates. We will follow their conventions below.

Erdmenger et al. \cite{Erdmenger:2002ex} change  basis to one in terms of 3d Grassmann spinors, in which it is possible to identify the $\NN_{\rm{3d}}=2$ content of $\NN_{\rm{4d}}=2$ superfields. Following \cite{Erdmenger:2002ex}, we term these coordinates $\theta,\stheta$, each of which is a 3d spinor with spinor index $\talpha = 1,2$.  They are obtained from the WB coordinates by first defining 
\begin{align}\label{WBtoEGKcoord}
\ttheta_{A}^{\talpha} =&~ \half ( \delta^{\talpha}_{~\alpha} \theta_{A}^{\alpha} + \delta^{\talpha}_{~\alphadot} \epsilon^{\alphadot\betadot} \thetabar_{\betadot}^{A})~, \cr
\tstheta_{A}^{\talpha} =&~ \frac{1}{2i} ( \delta^{\talpha}_{~\alpha} \theta_{A}^{\alpha} - \delta^{\talpha}_{~\alphadot} \epsilon^{\alphadot\betadot} \thetabar_{\betadot}^{A} )~.
\end{align}
 and then setting
 \begin{equation}\label{WBtoEGK2}
 \theta = \ttheta_1 - i \ttheta_2 ~, \qquad \stheta = \tstheta_1 - i \tstheta_2 ~.
 \end{equation}
 We denote these the EGK coordinates. Clearly \eqref{WBtoEGKcoord}, \eqref{WBtoEGK2} are not $\mathrm{SU}(2)_{\rm R}$-covariant. They also utilize isomorphisms from the 4d Weyl representations to the 3d Dirac representation. These isomorphisms respect a $\mathfrak{so}(1,2)$ subalgebra of the 4d Lorentz algebra. 
 
Setting $\stheta=0$ and $x^2=0$ isolates $\NN_{\rm 3d}=2$ superspace with coordinates $(x^\muhat, \theta,\thetabar)$ as a subsuperspace of $\NN_{\rm 4d}=2$ superspace. 
 
 It will be convenient below to write the change of coordinates and transformations among spinor fields as a matrix multiplication. We define
\begin{align}
& \theta_{\rm WB}^{T} = \left( \begin{array}{c c c c} \theta_{1}^{\alpha} & \thetabar^{1}_{\alphadot} & \theta_{2}^{\alpha} & \thetabar^{2}_{\alphadot} \end{array} \right)~, \qquad \theta_{\rm EGK}^T = \left( \begin{array}{c c c c} \theta^{\talpha} & \thetabar_{\talpha} & \stheta^{\talpha} & \sthetabar_{\talpha} \end{array} \right)~,
\end{align}
and
\begin{align}
\theta_{\rm WB}= \left( \begin{array}{c} \theta_{\alpha}^1 \\[1ex] \thetabar_{1}^{\alphadot} \\[1ex] \theta^{2}_{\alpha} \\[1ex] \thetabar_{2}^{\alphadot} \end{array} \right)~, \qquad  \theta_{\rm EGK} = \left( \begin{array}{c} \theta_{\talpha} \\[1ex] \thetabar^{\talpha} \\[1ex] \stheta_{\talpha} \\[1ex] \sthetabar^{\talpha} \end{array} \right) ~.
\end{align}
Then \eqref{WBtoEGKcoord}, \eqref{WBtoEGK2} take the form
\begin{equation}\label{eq:thetabasistrans}
\theta_{\rm EGK}^T = \theta_{\rm WB}^T S~\qquad\text{and}\qquad \theta_{\rm EGK} = S' \theta_{\rm WB} 
\end{equation}
with
\begin{equation}\label{Sdef}
S := \half \left( \begin{array}{c c c c} \delta^{~\talpha}_{\alpha}  & - \epsilon_{\alpha\talpha} & \delta_{\alpha}^{~\talpha}  & \epsilon_{\alpha\talpha} \\[1ex] - \epsilon^{\alphadot\talpha} & \delta^{\alphadot}_{~\talpha} & \epsilon^{\alphadot\talpha} & \delta^{\alphadot}_{~\talpha} \\[1ex] -i \delta^{~\talpha}_{\alpha} & -i \epsilon_{\alpha\talpha} & -i \delta_{\alpha}^{~\talpha} & i \epsilon_{\alpha\talpha} \\[1ex]  i \epsilon^{\alphadot\talpha} & i \delta^{\alphadot}_{~\talpha} & -i \epsilon^{\alphadot\talpha} & i \delta^{\alphadot}_{~\talpha} \end{array} \right)%
\qquad \text{and} \qquad
 S' :=  \half \left(\begin{array}{c c c c} -i \delta_{\talpha}^{~\alpha} & -i \epsilon_{\talpha\alphadot} & -\delta_{\talpha}^{~\alpha} & - \epsilon_{\talpha\alphadot}  \\[1ex] i \epsilon^{\talpha\alpha} & i \delta^{\talpha}_{~\alphadot} & - \epsilon^{\talpha\alpha} & - \delta^{\talpha}_{~\alphadot} \\[1ex] -i \delta_{\talpha}^{~\alpha} & i \epsilon_{\talpha\alphadot} & -\delta_{\talpha}^{~\alpha} & \epsilon_{\talpha\alphadot} \\[1ex] -i \epsilon^{\talpha\alpha} &  i \delta^{\talpha}_{~\alphadot} & \epsilon^{\talpha\alpha} & - \delta^{\talpha}_{~\alphadot} \end{array} \right)~.
\end{equation}
Since $\theta,\thetabar$ transform in the same representation of $\mathfrak{so}(1,2)$ they can be contracted.  The standard convention for this contraction is $\theta \thetabar = \theta^{\talpha} \thetabar_{\talpha}$.  As usual, $\theta \theta = \theta^{\talpha} \theta_{\talpha}$ and $\thetabar \thetabar = \thetabar_{\talpha} \thetabar^{\talpha}$.

\subsection{Mapping Between Superspace-Based Formulations}
The field content in the WB basis consists of an
$\NN_{\rm 4d} = 2$ vector multiplet packaged into a ($\mathfrak{g}_{\mathbbm{C}}$-valued) chiral object, $\Psi$, which can be expanded in terms of ($\mathfrak{g}_{\mathbbm{C}}$-valued) $\NN_{\rm 4d} = 1$ superfields as 
\begin{equation}\label{PsiN1exp}
\Psi(\tilde{y}, \theta_A) = \Phi(\tilde{y}, \theta_1) + i \sqrt{2} \, \theta_{2}^\alpha W_{\alpha}(\tilde{y}, \theta_1) + \theta_{2}^\alpha \theta_{2\alpha} G(\tilde{y},\theta_1)~.
\end{equation}
where $\tilde{y}$ is the $\NN=2$ version of the chiral coordinate,
\begin{equation}
\tilde{y}^\mu := x^\mu + i \theta_{A}^{\alpha} \sigma^{\mu}_{\alpha\betadot} \thetabar^{A\betadot} ~.
\end{equation}
In Wess-Zumino gauge, the $\NN_{\rm{4d}}=1$ superfields are given by
\begin{align}\label{4dFields}
\Phi(\tilde{y},\theta_1) =&~ \phi(\tilde{y}) + \sqrt{2} \theta_{1} \psi(\tilde{y}) + \theta_1 \theta_{1} F(\tilde{y}) ~,  \cr
W_{\alpha}(\tilde{y},\theta_1) =&~ -i \xi_{\alpha}(\tilde{y}) + \left[ \delta_{\alpha}^{\phantom{\alpha}\beta} D(\tilde{y})  -i (\sigma^{\mu\nu})_{\alpha}^{\phantom{\alpha}\beta} F_{\mu\nu}(\tilde{y}) \right] \theta_{1\beta} + \theta_{1} \theta_{1} \sigma_{\alpha\alphadot}^{\mu} \DD_\mu \xibar^{\alphadot}(\tilde{y})~\cr
G(\tilde{y},\theta_1) =&~  \Fbar(\tilde{y}) + 2 \theta_1 [\xi(\tilde{y}), \phibar(\tilde{y})] + i \sqrt{2} \theta_1 \sigma^\mu \DD_\mu \psibar(\tilde{y}) + \cr
&~  + \theta_1 \theta_1 \left( i [D(\tilde{y}),\phibar(\tilde{y})] + \sqrt{2} [\xibar(\tilde{y}),\psibar(\tilde{y})] + \DD^\mu \DD_\mu \phibar(\tilde{y}) \right) ~,
\end{align}
where $F_{\mu\nu} = 2 \partial_{[\mu} A_{\nu]} + [A_\mu, A_\nu]$.  Here $W_{\alpha}(y,\theta)$ is obtained from the vector superfield 
\begin{equation}
V(y,\theta) = - \theta \sigma^\mu \thetabar A_\mu(y) + i (\theta\theta) \thetabar \xibar(y) - i (\thetabar\thetabar) \theta \xi(y) + \half \theta \theta \thetabar \thetabar ( D(y) - i \pd^\mu A_\mu(y) )~,
\end{equation}
as in \cite{Wess:1992cp}.

The $\NN_{\rm 4d}=2$ hypermultiplet, transforming in the adjoint representation of the gauge group, consists of an $\mathrm{SU}(2)_R$ doublet of complex scalars, an  $\mathrm{SU}(2)_R$ doublet of auxiliary scalars, and a single,  $\mathrm{SU}(2)_R$-singlet Dirac fermion. (See e.g. \cite{Sohnius:1985qm}.) Breaking the Dirac fermion into its chiral and anti-chiral components, one can group these fields into two $\NN_{\rm 4d}=1$ chiral multiplets, which is how the $\NN_{\rm 4d}=2$ hypermultiplet is conventionally presented in the literature. We define this set of chiral superfields as
\begin{equation}
\QQ_{1,2} = Q_{1,2}(y) + \sqrt{2} \theta \psi_{1,2}(y) + F_{1,2}(y)~,
\end{equation}
where now $y$ is the conventional $\NN_{4d}=1$ chiral coordinate. Note that the indices on this pair are not $\mathrm{SU}(2)_{\rm R}$ indices.

The $\NN_{\rm 4d} = 4$ action in terms of these superfields is
\begin{align}\label{4dN1action}
S_{\rm amb} =&~ \frac{1}{g_{\rm ym}^2} \Im \left[ i \int d^4 x \left( \int d^2\theta \Tr W^{\alpha} W_{\alpha} + 2 \int d^2 \theta d^2 \thetabar \Tr \left\{ e^{2iV} \Phibar e^{-2iV} \Phi \right\} \right) \right]   + \cr
&~ + \frac{2}{g_{\rm ym}^2} \int d^4 x  \bigg\{ \int d^4 \theta \Tr \left( e^{2i V} \QQbar_{1} e^{-2i V} \QQ_{1} + e^{2i V} \QQbar_{2} e^{-2i V}  \QQ_2 \right) + \cr
&~ \qquad \qquad \qquad \qquad +  \int d^2 \theta \Tr \left( \sqrt{2} \QQ_{2} [\Phi, \QQ_1]   + \textrm{~c.c.~} \right)  \bigg\}~.
\end{align}

Because the fermions mix in a nontrivial way, it will be convenient to define 4-vectors of the fermions as
\begin{align}\label{eq:fermionWBdef}
\xi_{\rm WB} :=
 \left( \begin{array}{c} \psi_{\alpha} \\[1ex] \psibar^{\alphadot} \\[1ex] \xi_{\alpha} \\[1ex] \xibar^{\alphadot} \end{array} \right) \qquad\text{and}\qquad 
\psi_{\rm WB} := \left( \begin{array}{c} \psi_{1\alpha} \\[1ex] \psibar_{1}^{\alphadot} \\[1ex] \psi_{2\alpha} \\[1ex] \psibar_{2}^{\alphadot} \end{array} \right)~.
\end{align}
In the EGK basis, meanwhile,  the $\NN_{\rm 4d}=2$ vector field content can be regrouped into the component fields of an $\NN_{\rm 3d}=2$ vector multiplet $\VV$ and chiral multiplets $\upphi$ and $\overbar{\upphi}$ parameterized as
\begin{align}
\VV (\yhat,\theta)=&~ -i \theta\thetabar \rho(\yhat) - \theta \sigma^{\muhat} \thetabar v_{\muhat}(\yhat) + i (\theta\theta) \thetabar \bupxibar(\yhat) - i (\thetabar\thetabar) \theta\bupxi(\yhat) + \half \theta \theta \thetabar \thetabar \left( d(\yhat) - i\pd^{\muhat} v_{\muhat}(\yhat) \right)~\cr
\upphi (\yhat,\theta)=&~ \varphi(\yhat;s) + \sqrt{2} \theta \bpsi(\yhat;s) + \theta\theta f(\yhat;s) ~\raisetag{20pt}
\end{align}
written according to the 3d version of  the chiral coordinates
\begin{equation}
\yhat^{\muhat} := \xhat^{\muhat} + i \theta \sigma^{\muhat} \thetabar ~, \qquad \yhatbar^{\muhat} = \xhat^{\muhat} - i \theta \sigma^{\muhat} \thetabar ~.
\end{equation}
The $\NN_{\rm 4d}=2$ hypermultiplet, on the other hand, decomposes into a doublet of $\NN_{\rm 3d} = 2$ chiral superfields.
Here $\UU^{1,2}$ is a pair of $\NN_{\rm 3d}=2$ chiral superfields with expansions 
\begin{equation}\label{UUchiral}
\UU^{1,2}(\yhat) = w^{1,2}(\yhat) + \sqrt{2} \theta \bchi^{1,2}(\yhat) + \theta^2 f^{1,2}(\yhat) ~.
\end{equation}
The conjugate anti-chiral superfields are denoted $\UUbar_{1,2}$.  In order to construct the $\NN_{\rm 4d} =4$ ambient action in terms these superfields one needs the linear multiplet
 \begin{align}
 \Sigma :=&~ \epsilon^{\talpha\tbeta} \Dbar_{\talpha} (e^{2i \VV} D_{\tbeta} e^{-2i \VV})  \cr
 =&~  4 \rho(\yhat) + 4 \thetabar \bupxi(\yhat) -4 \theta \bupxibar(\yhat) + 4i \theta\thetabar D(\yhat) - 4i \theta \sigma^{\muhat} \thetabar \DD_{\muhat} \rho(\yhat) - 2 \epsilon^{\muhat\nuhat\kappahat} \theta \sigma_{\kappahat} \thetabar F_{\muhat\nuhat}(\yhat) + \cr
&~ + 4 \theta \theta [\rho(\yhat), \thetabar \bupxibar(\yhat)] + 4i (\theta\theta) \thetabar \sigma^{\muhat} \DD_{\muhat} \bupxibar(\yhat)
 \end{align}
 where the supercovariant derivatives with respect to the 3d chiral coordinates are
\begin{align}\label{DDbar3dchiral}
D_\talpha =&~ \frac{\pd}{\pd \theta^\talpha} + 2i \sigma^{\muhat}_{\talpha\tbeta} \thetabar^\tbeta \frac{\pd}{\pd \yhat^{\muhat}} ~, \qquad \Dbar_{\talpha} = - \frac{\pd}{\pd \thetabar^{\talpha}} ~.
\end{align}
The ambient $\NN_{\rm 4d} = 4$ action in terms of these superfields takes the form \cite{Cherkis:2011ee}
\begin{align}\label{3DN2action}
S_{\rm amb} :=&~ \frac{1}{2 g_{\rm ym}^2} \int d^4x  \Tr \bigg\{ \int d^2 \theta d^2\thetabar \bigg[ \left( \upphi + e^{2i \VV} \upphibar e^{-2i\VV} - i e^{2i \VV} \pd_2 e^{-2i\VV} \right)^2 - \frac{1}{4} \Sigma^2  + \cr
&~ \qquad \qquad \qquad \qquad \qquad \qquad   +  2 \left( e^{2i \VV} \UUbar_1 e^{-2i \VV} \UU^1 + e^{2i \VV} \UUbar_2 e^{-2i \VV} \UU^2  \right)      \bigg]  + \cr
&~ \qquad  \qquad \qquad  \quad +  \int d^2 \theta \left( \UU^1 [ i\pd_2 + \upphi, \UU^2] -  \UU^2 [ i\pd_2 + \upphi, \UU^1] \right)  ~+~\text{~c.c.}    \bigg\} ~. \qquad
\end{align}

As above, we group the fermions into 4-vectors as
\begin{align}\label{eq:fermionEGKdef}
\boldsymbol{\upxi}_{\rm EGK} :=\left(\begin{array}{c} \frac{1}{\sqrt{2}} \bpsi_{\talpha} \\[1ex] \frac{1}{\sqrt{2}} \bpsibar^{\talpha} \\[1ex] i \bupxibar_{\talpha} \\[1ex] -i \bupxi^{\talpha} \end{array} \right) \qquad \text{and}\qquad
\bchi_{\rm EGK} := \left( \begin{array}{c} i \bchibar_{2\talpha} \\[1ex] -i \bchi^{2\talpha} \\[1ex] \bchi_{\talpha}^1 \\[1ex] \bchibar_{1}^{\talpha} \end{array} \right)~.
\end{align}
The factors of $i$ and $\sqrt{2}$ in the various components of these 4-vectors are present so that the map can be expressed simply in terms of the $S$ and $S'$ matrices \eqref{Sdef}. One could remove them by modifying the form of the superfield expansions, but we have instead chosen to follow standard conventions regarding these.

By requiring that the ambient actions, \eqref{4dN1action} and \eqref{3DN2action}, are the same, one obtains the map relating the EGK and WB formulations:
\begin{align}\label{na4d3dmap}
& \phi = \frac{1}{\sqrt{2}} (\Re{\varphi} - i\rho)~,\qquad A_{\mu} = (v_{\muhat}, \Im{\varphi})~, \cr
& F = \frac{1}{\sqrt{2}} (d + \DD_2 (\Re{\varphi})) + \frac{1}{2\sqrt{2}} (f - \fbar) ~,\qquad D - i [\phi,\phibar] = -\half (f+ \fbar)~,
&\cr
&\left( \begin{array}{c} \Re{Q_1} \\[1ex] \Im{Q_1} \\[1ex] \Re{Q_2} \\[1ex] \Im{Q_2} \end{array} \right) = \frac{1}{\sqrt{2}} \left( \begin{array}{c c c c} 0 & 0 & 0 & -1 \\[1ex] 0 & 1 & 0 & 0 \\[1ex] 0 & 0 & -1 & 0 \\[1ex] -1 & 0 & 0 & 0 \end{array} \right) \left( \begin{array}{c} \Re{w^1} \\[1ex] \Im{w^1} \\[1ex] \Re{w^2} \\[1ex] \Im{w^2} \end{array} \right) \cr
&\psi_{\rm WB}=\frac{1}{\sqrt{2}}S \bchi_{\rm EGK}~, \qquad \xi_{\rm WB} = S\boldsymbol{\upxi}_{\rm EGK} 
\end{align}
with the matrix $S$ as given in \eqref{Sdef}.   This is the nonabelian version of the  map given in  \cite{Erdmenger:2002ex}'s equations (2.18)-(2.21).

\subsection{Map to $\mathrm{SU}(2)\times \mathrm{SU}(2)$-invariant Form}
We now provide the map from the WB formulation to the formulation manifesting $\mathrm{SU}(2)_{\rm V}\times \mathrm{SU}(2)_{\rm H}$ R-symmetry used in this work.

The gauge fields and the $D$ auxiliary are unchanged, and the complex scalars from the WB language group into triplets  indexed by $r=1,2,3$:
\begin{align}
X_r := X^{\rm{H}}_r + i X^{\rm{V}}_\rtilde := (Q_1,~ Q_2, ~\phi)~\qquad F_r:= (F_1, F_2, F) ~.
\end{align}
Note that the definition of $X^{\rm{H}}_r$ and $X^{\rm{V}}_\rtilde$ as the real and imaginary parts of $X_r$, respectively, is exactly swapped compared to the definition of \cite{DeWolfe:2001pq}.
For the fermions, we first define Majorana spinors in the usual Weyl basis as
\begin{align}
\xi^{\rm M} = \left( \begin{array}{c} \xi_\alpha \\ \xibar^{\alphadot} \end{array} \right) ~, \qquad \psi^{\rm M} = \left( \begin{array}{c} \psi_{\alpha} \\ \psibar^{\alphadot} \end{array} \right) ~,\qquad \text{etc.}
\end{align}
We can then group the four fermion fields from the $\NN_{\text{4d}}=2$ vector and hyper into an $\mathrm{SO}(4)_{\rm R}$ quartet as $\lambda_I := (\psi_1^{\rm M}, \ \psi_2^{\rm M}, \ \psi^{\rm M}, \ \xi^{\rm M})$ with $I=1,\dots 4$.

%

\begin{thebibliography}{10}

\bibitem{Hanany:1996ie}
A.~Hanany and E.~Witten, ``{Type IIB superstrings, BPS monopoles, and
  three-dimensional gauge dynamics},''
  \href{http://dx.doi.org/10.1016/S0550-3213(97)00157-0,
  10.1016/S0550-3213(97)80030-2}{{\em Nucl. Phys.} {\bfseries B492} (1997)
  152--190},
\href{http://arxiv.org/abs/hep-th/9611230}{{\ttfamily arXiv:hep-th/9611230
  [hep-th]}}.

\bibitem{Cherkis:1997aa}
S.~A. Cherkis and A.~Kapustin, ``{Singular monopoles and supersymmetric gauge
  theories in three-dimensions},''
  \href{http://dx.doi.org/10.1016/S0550-3213(98)00341-1}{{\em Nucl.Phys.}
  {\bfseries B525} (1998) 215--234},
\href{http://arxiv.org/abs/hep-th/9711145}{{\ttfamily arXiv:hep-th/9711145
  [hep-th]}}.

\bibitem{Karch:2000gx}
A.~Karch and L.~Randall, ``{Open and closed string interpretation of SUSY CFT's
  on branes with boundaries},''
  \href{http://dx.doi.org/10.1088/1126-6708/2001/06/063}{{\em JHEP} {\bfseries
  06} (2001) 063},
\href{http://arxiv.org/abs/hep-th/0105132}{{\ttfamily arXiv:hep-th/0105132
  [hep-th]}}.

\bibitem{DeWolfe:2001pq}
O.~DeWolfe, D.~Z. Freedman, and H.~Ooguri, ``{Holography and defect conformal
  field theories},'' \href{http://dx.doi.org/10.1103/PhysRevD.66.025009}{{\em
  Phys.Rev.} {\bfseries D66} (2002) 025009},
\href{http://arxiv.org/abs/hep-th/0111135}{{\ttfamily arXiv:hep-th/0111135
  [hep-th]}}.

\bibitem{Erdmenger:2002ex}
J.~Erdmenger, Z.~Guralnik, and I.~Kirsch, ``{Four-dimensional superconformal
  theories with interacting boundaries or defects},''
  \href{http://dx.doi.org/10.1103/PhysRevD.66.025020}{{\em Phys. Rev.}
  {\bfseries D66} (2002) 025020},
\href{http://arxiv.org/abs/hep-th/0203020}{{\ttfamily arXiv:hep-th/0203020
  [hep-th]}}.

\bibitem{deLeeuw:2015hxa}
M.~de~Leeuw, C.~Kristjansen, and K.~Zarembo, ``{One-point Functions in Defect
  CFT and Integrability},''
  \href{http://dx.doi.org/10.1007/JHEP08(2015)098}{{\em JHEP} {\bfseries 08}
  (2015) 098}, \href{http://arxiv.org/abs/1506.06958}{{\ttfamily
  arXiv:1506.06958 [hep-th]}}.

\bibitem{DeLeeuw:2018cal}
M.~De~Leeuw, C.~Kristjansen, and G.~Linardopoulos, ``{Scalar one-point
  functions and matrix product states of AdS/dCFT},''
  \href{http://dx.doi.org/10.1016/j.physletb.2018.03.083}{{\em Phys. Lett. B}
  {\bfseries 781} (2018) 238--243},
  \href{http://arxiv.org/abs/1802.01598}{{\ttfamily arXiv:1802.01598
  [hep-th]}}.

\bibitem{Komatsu:2020sup}
S.~Komatsu and Y.~Wang, ``{Non-perturbative defect one-point functions in
  planar $\mathcal{N}=4$ super-Yang-Mills},''
  \href{http://dx.doi.org/10.1016/j.nuclphysb.2020.115120}{{\em Nucl. Phys. B}
  {\bfseries 958} (2020) 115120},
  \href{http://arxiv.org/abs/2004.09514}{{\ttfamily arXiv:2004.09514
  [hep-th]}}.

\bibitem{Gaiotto:2008sa}
D.~Gaiotto and E.~Witten, ``{Supersymmetric Boundary Conditions in N=4 Super
  Yang-Mills Theory},'' \href{http://dx.doi.org/10.1007/s10955-009-9687-3}{{\em
  J. Statist. Phys.} {\bfseries 135} (2009) 789--855},
\href{http://arxiv.org/abs/0804.2902}{{\ttfamily arXiv:0804.2902 [hep-th]}}.

\bibitem{Jensen:2010ga}
K.~Jensen, A.~Karch, D.~T. Son, and E.~G. Thompson, ``{Holographic
  Berezinskii-Kosterlitz-Thouless Transitions},''
  \href{http://dx.doi.org/10.1103/PhysRevLett.105.041601}{{\em Phys. Rev.
  Lett.} {\bfseries 105} (2010) 041601},
  \href{http://arxiv.org/abs/1002.3159}{{\ttfamily arXiv:1002.3159 [hep-th]}}.

\bibitem{Witten:1988ze}
E.~Witten, ``{Topological Quantum Field Theory},''
  \href{http://dx.doi.org/10.1007/BF01223371}{{\em Commun. Math. Phys.}
  {\bfseries 117} (1988) 353}.

\bibitem{Cherkis:2011ee}
S.~A. Cherkis, C.~O'Hara, and C.~Saemann, ``{Super Yang-Mills Theory with
  Impurity Walls and Instanton Moduli Spaces},''
  \href{http://dx.doi.org/10.1103/PhysRevD.83.126009}{{\em Phys. Rev.}
  {\bfseries D83} (2011) 126009},
\href{http://arxiv.org/abs/1103.0042}{{\ttfamily arXiv:1103.0042 [hep-th]}}.

\bibitem{Witten:1978mh}
E.~Witten and D.~I. Olive, ``{Supersymmetry Algebras That Include Topological
  Charges},''
\href{http://dx.doi.org/10.1016/0370-2693(78)90357-X}{{\em Phys. Lett.}
  {\bfseries B78} (1978) 97--101}.

\bibitem{Kapustin:2006pk}
A.~Kapustin and E.~Witten, ``{Electric-Magnetic Duality And The Geometric
  Langlands Program},''
  \href{http://dx.doi.org/10.4310/CNTP.2007.v1.n1.a1}{{\em Commun. Num. Theor.
  Phys.} {\bfseries 1} (2007) 1--236},
\href{http://arxiv.org/abs/hep-th/0604151}{{\ttfamily arXiv:hep-th/0604151
  [hep-th]}}.

\bibitem{Witten:2011zz}
E.~Witten, ``Fivebranes and knots,'' {\em Quantum Topol.} {\bfseries 3} no.~1,
  (2012) 1--137, \href{http://arxiv.org/abs/1101.3216}{{\ttfamily
  arXiv:1101.3216 [hep-th]}}.

\bibitem{Gaiotto:2011nm}
D.~Gaiotto and E.~Witten, ``{Knot Invariants from Four-Dimensional Gauge
  Theory},'' \href{http://dx.doi.org/10.4310/ATMP.2012.v16.n3.a5}{{\em Adv.
  Theor. Math. Phys.} {\bfseries 16} no.~3, (2012) 935--1086},
\href{http://arxiv.org/abs/1106.4789}{{\ttfamily arXiv:1106.4789 [hep-th]}}.

\bibitem{Domokos:2017wlb}
S.~K. Domokos and A.~B. Royston, ``{Holography for field theory solitons},''
  \href{http://dx.doi.org/10.1007/JHEP07(2017)065}{{\em JHEP} {\bfseries 07}
  (2017) 065},
\href{http://arxiv.org/abs/1706.00425}{{\ttfamily arXiv:1706.00425 [hep-th]}}.

\bibitem{Wess:1992cp}
J.~Wess and J.~Bagger, {\em {Supersymmetry and supergravity}}.
\newblock Princeton University Press, Princeton, NJ, USA, 1992.

\bibitem{Gates:1983nr}
S.~J. Gates, M.~T. Grisaru, M.~Rocek, and W.~Siegel, {\em {Superspace Or One
  Thousand and One Lessons in Supersymmetry}}, vol.~58 of {\em Frontiers in
  Physics}.
\newblock 1983.
\newblock \href{http://arxiv.org/abs/hep-th/0108200}{{\ttfamily
  arXiv:hep-th/0108200}}.

\bibitem{Gaiotto:2008sd}
D.~Gaiotto and E.~Witten, ``{Janus Configurations, Chern-Simons Couplings, And
  The theta-Angle in N=4 Super Yang-Mills Theory},''
  \href{http://dx.doi.org/10.1007/JHEP06(2010)097}{{\em JHEP} {\bfseries 06}
  (2010) 097}, \href{http://arxiv.org/abs/0804.2907}{{\ttfamily arXiv:0804.2907
  [hep-th]}}.

\bibitem{Sohnius:1985qm}
M.~F. Sohnius, ``{Introducing Supersymmetry},''
  \href{http://dx.doi.org/10.1016/0370-1573(85)90023-7}{{\em Phys. Rept.}
  {\bfseries 128} (1985) 39--204}.

\bibitem{Sethi:1997zza}
S.~Sethi, ``{The Matrix formulation of type IIb five-branes},''
  \href{http://dx.doi.org/10.1016/S0550-3213(98)00302-2}{{\em Nucl. Phys. B}
  {\bfseries 523} (1998) 158--170},
  \href{http://arxiv.org/abs/hep-th/9710005}{{\ttfamily arXiv:hep-th/9710005}}.

\bibitem{Kapustin:1998pb}
A.~Kapustin and S.~Sethi, ``{The Higgs branch of impurity theories},''
  \href{http://dx.doi.org/10.4310/ATMP.1998.v2.n3.a6}{{\em Adv. Theor. Math.
  Phys.} {\bfseries 2} (1998) 571--591},
\href{http://arxiv.org/abs/hep-th/9804027}{{\ttfamily arXiv:hep-th/9804027
  [hep-th]}}.

\bibitem{MR3289327}
R.~Mazzeo and E.~Witten, ``The {N}ahm pole boundary condition,'' in {\em The
  influence of {S}olomon {L}efschetz in geometry and topology}, vol.~621 of
  {\em Contemp. Math.}, pp.~171--226.
\newblock Amer. Math. Soc., Providence, RI, 2014.
\newblock \href{http://arxiv.org/abs/1311.3167}{{\ttfamily arXiv:1311.3167
  [math.DG]}}.

\bibitem{Mazzeo:2017qwz}
R.~Mazzeo and E.~Witten, ``{The KW equations and the Nahm pole boundary
  condition with knots},''
  \href{http://dx.doi.org/10.4310/CAG.2020.v28.n4.a4}{{\em Commun. Anal. Geom.}
  {\bfseries 28} no.~4, (2020) 871--942},
  \href{http://arxiv.org/abs/1712.00835}{{\ttfamily arXiv:1712.00835
  [math.DG]}}.

\bibitem{MR4019897}
S.~He and R.~Mazzeo, ``The extended {B}ogomolny equations and generalized
  {N}ahm pole boundary condition,'' {\em Geom. Topol.} {\bfseries 23} no.~5,
  (2019) 2475--2517, \href{http://arxiv.org/abs/1710.10645}{{\ttfamily
  arXiv:1710.10645 [math.DG]}}.

\bibitem{MR4139043}
S.~He and R.~Mazzeo, ``The extended {B}ogomolny equations with generalized
  {N}ahm pole boundary conditions, {II},'' {\em Duke Math. J.} {\bfseries 169}
  no.~12, (2020) 2281--2335, \href{http://arxiv.org/abs/1806.06314}{{\ttfamily
  arXiv:1806.06314 [math.DG]}}.

\bibitem{Moore:2015szp}
G.~W. Moore, A.~B. Royston, and D.~Van~den Bleeken, ``{Semiclassical framed BPS
  states},'' \href{http://dx.doi.org/10.1007/JHEP07(2016)071}{{\em JHEP}
  {\bfseries 07} (2016) 071}, \href{http://arxiv.org/abs/1512.08924}{{\ttfamily
  arXiv:1512.08924 [hep-th]}}.

\bibitem{Moore:2015qyu}
G.~W. Moore, A.~B. Royston, and D.~Van~den Bleeken, ``{$L^2$-Kernels Of
  Dirac-Type Operators On Monopole Moduli Spaces},'' {\em Proc. Symp. Pure
  Math.} (2015) 169--182, \href{http://arxiv.org/abs/1512.08923}{{\ttfamily
  arXiv:1512.08923 [hep-th]}}.

\bibitem{Brennan:2016znk}
T.~D. Brennan and G.~W. Moore, ``{A note on the semiclassical formulation of
  BPS states in four-dimensional $N=$ 2 theories},''
  \href{http://dx.doi.org/10.1093/ptep/ptw159}{{\em PTEP} {\bfseries 2016}
  no.~12, (2016) 12C110}, \href{http://arxiv.org/abs/1610.00697}{{\ttfamily
  arXiv:1610.00697 [hep-th]}}.

\bibitem{Brennan:2018ura}
T.~D. Brennan, G.~W. Moore, and A.~B. Royston, ``{Wall Crossing from Dirac
  Zeromodes},'' \href{http://dx.doi.org/10.1007/JHEP09(2018)038}{{\em JHEP}
  {\bfseries 09} (2018) 038}, \href{http://arxiv.org/abs/1805.08783}{{\ttfamily
  arXiv:1805.08783 [hep-th]}}.

\bibitem{MooreTalk}
D.~Gaiotto, A.~Kahn, G.~Moore, and F.~Yan, ``{2d Categorical Wall-Crossing With
  Twisted Masses, And An Application To Knot Invariants},''
\newblock 2 Jun 2021.
\newblock Presented by Gregory Moore at Number Theory, Strings, and Quantum
  Physics at IPMU.

\end{thebibliography}
%

\providecommand{\href}[2]{#2}\begingroup\raggedright\endgroup

\end{document}